\documentclass[prl,twocolumn,showkeys,showpacs,superscriptaddress,10pt
]{revtex4-1}
\usepackage{graphicx}
\usepackage{grffile}
\usepackage{amsthm}
\usepackage{epsfig}
\usepackage{amsmath,amssymb,amsfonts,mathtools,bbm,MnSymbol,mathrsfs,xfrac}
\usepackage{longtable}
\usepackage{tabularx}
\usepackage{colortbl}
\usepackage[table]{xcolor}
\usepackage[breaklinks=true,colorlinks=true,linkcolor=blue,urlcolor=blue,citecolor=blue]{hyperref}
\usepackage{comment}
\usepackage{color}
\usepackage[makeroom]{cancel}

\usepackage{soul}
\pdfoutput=1

\newcommand{\ent}{{\mathrm{ent}}}
\renewcommand{\[}{\begin{equation}}
\renewcommand{\]}{\end{equation}}

\newcommand{\ket}[1]{|#1\rangle}

\newcommand{\pro}[2]{|#1\rangle\langle#2|}

\newcommand{\abs}[1]{|#1|}
\newcommand{\ov}[1]{\overline{#1}}
\newcommand{\tr}{\mathrm{tr}}

\newcommand{\R}{{\hat{\rho}}}

\newcommand{\HS}{\mathcal{H}}

\definecolor{mygray}{gray}{0.6}

\newcommand{\na}{{n_{\!A}}}

\theoremstyle{definition}

\definecolor{dfcol}{cmyk}{1, 0.2108, 0.13, 0.3}
\newcommand{\df}[1]{\ifthenelse{\boolean{}}{\textcolor{dfcol}{[{\bf DF}: #1]}}{}}
\begin{document}

\title{How much entanglement can be created in a closed system?}

\author{Dana Faiez}
\email{dfaiez@ucsc.edu}
\affiliation{Department of Physics, University of California, Santa Cruz, California 95064, USA}
\author{Dominik \v{S}afr\'{a}nek}
\email{dsafrane@ucsc.edu}
\affiliation{SCIPP and Department of Physics, University of California, Santa Cruz, California 95064, USA}

\date{\today}

\begin{abstract}
In a closed system, the total number of particles is fixed. We ask how much does this conservation law restrict the amount of entanglement that can be created.  We derive a tight upper bound on the bipartite entanglement entropy in closed systems, and find what a maximally entangled state looks like in such a system.
Finally, we illustrate numerically on an isolated system of one-dimensional fermionic gas, that the upper bound can be reached during its unitary evolution, when starting in a pure state that emulates a thermal state with high enough temperature. These results are in accordance with current experiments measuring R\'enyi-2 entanglement entropy, all of which employ a particle-conserving Hamiltonian, where our bound acts as a loose bound, and will become especially important for bounding the amount of entanglement that can be spontaneously created, once a direct measurement of entanglement entropy becomes feasible.
\end{abstract}

\maketitle
Entanglement is one of the most intriguing characteristics of quantum systems.
It evolved from its perception as a mathematical artifact, as a result of EPR paradox~\cite{nikolic2012epr}, to becoming closely related and applicable to the fields of condensed matter~\cite{osborne2002entanglement,amico2008entanglement, bloch2008many,lee2009lattice,  zhang2011entanglement,jiang2012identifying}, quantum information~\cite{ekert1991quantum,bennett1993teleporting,nielsen2002quantum,wendin2017quantum, ng2018resource,chitambar2019quantum,landsberg2019very}, quantum metrology~\cite{giovannetti2006quantum, dowling2008quantum,giovannetti2011advances,demkowicz2012elusive, toth2014quantum,szczykulska2016multi}, and quantum gravity~\cite{dewitt2008introduction,takayanagi2012entanglement,dong2014holographic,schulz2014review,nishioka2018entanglement}.

In the field of quantum information, entangled states are the backbone of quantum information protocols as they are considered a resource for tasks such as quantum
teleportation~\cite{bennett1993teleporting,helwig2012absolute},  cryptography~\cite{ekert1991quantum}, and dense coding~\cite{bennett1992communication}.

In these quantum information protocols, more entanglement usually leads to a better performance. Therefore, it is important to set precise upper bounds on how much entanglement is in principle available in performing these tasks~\cite{zozulya2007bipartite,li2010measurable,jevtic2012maximally,avery2014universal,song2016lower,vidmar2017entanglement,vidmar2017entanglementquadratic,beaud2018bounds,bauml2018fundamental,fujita2018page,huang2019universal,huang2019dynamics}.

As different tasks require different types of entangled states, numerous measures of entanglement have been introduced~\cite{wiseman2003entanglement, bennett2011postulates,plenio2014introduction,girolami2017quantifying}. The most prominent measure of entanglement is entanglement entropy~\cite{page1993average,vidmar2017entanglement,vidmar2017entanglementquadratic}. It is defined as the von Neumann entropy of the reduced density matrix $\R_A=\tr_B[\pro{\psi}{\psi}]$, where $\ket{\psi}$ denotes the state of the composite system,
\[
\label{eq:ententropy}
S_{\mathrm{ent}}\equiv S(\R_A).
\]

This is a valuable measure as it draws a direct connection between density matrix and the amount of non-local correlations present in a given system.
Entanglement entropy also gained significant attention in the past few decades due to the discovery of its geometric scaling in thermal state as well as ground states (famously known as the volume law~\cite{nakagawa2018universality} and the area law~\cite{eisert2008area,laflorencie2016quantum,cho2018realistic} respectively), and its use for characterizing quantum phase transition~\cite{osborne2002entanglement, gu2004entanglement, le2008entanglement, barghathi2019operationally}.

Despite its importance, this quantity has proven extremely difficult to probe experimentally, and related R\'enyi-2 entanglement entropy has been measured instead~\cite{islam2015measuring,kaufman2016quantum,brydges2018probing}.
However, an experimental proposal for measuring the entanglement entropy has been put forward recently~\cite{mendes2019measuring}, opening new exciting possibilities.

There exists a general bound on entanglement entropy. For a pure state of a bipartite system, it is straight forward to show that $S_{\mathrm{ent}}\equiv S(\R_A)=S(\R_B)$. This leads to~\cite{nishioka2018entanglement},
\[
\label{eq:originalbound}
S_{\mathrm{ent}}\leq \ln \min\Big\{\dim \HS_A,\dim \HS_B\Big\}.
\]

However, one could wonder whether Eq.~\eqref{eq:originalbound} is stringent enough for systems with additional conservation laws, that effectively restrict some degrees of freedom.

For example, consider a system of $2$ fermionic particles contained on a lattice comprising of 6 sites, partitioned into two sublattices of 3 sites. Since there can be any number $0$, $1$, and $2$ particles in each sublattice, the upper bound on entanglement entropy given by Eq.~\eqref{eq:originalbound} is $S_{\mathrm{ent}}\leq \ln\big(\binom{3}{0}+\binom{3}{1}+\binom{3}{2}\big)=\ln 7$, yet because of the conservation law, this could be considerably larger than the actually achievable entropy.

This is important because, among the aforementioned quantum tasks, those that incorporate massive particles --- such as the constituents of condensed matter systems --- often exhibit constrains such as the conservation of the total number of particles or charge~\cite{daley2012measuring,islam2015measuring,kaufman2016quantum,pichler2016measurement,barghathi2018r,linke2018measuring,brydges2018probing,lukin2019probing}. Such restrictions are described by superselection rules~\cite{bartlett2007reference,benatti2012bipartite}.
It has been suggested that these restrictions can in fact be used as a resource and can enhance the security of quantum communication~\cite{ionicioiu2002quantum,verstraete2003quantum,bartlett2007reference,marzolino2015quantum,marzolino2016performances} and measurement accuracy~\cite{benatti2010sub,gross2010nonlinear,benatti2011entanglement,benatti2014entanglement}.
However, among the vast literature on quantum information protocols, specific bounds on entanglement entropy in the presence of superselection rules are not sufficient.

Given the commonality of these conservation laws and recent efforts in probing entanglement entropy experimentally, it is an incentive to provide precise bounds for this quantity. In this Rapid Communication, we derive a general tight upper bound on entanglement entropy for closed systems (in thermodynamic sense), which are defined as those where the total number of particles stays constant. This can be applied to quantum systems evolved with a time-independent or a time-dependent Hamiltonian, as long as this evolution conserves the total number of particles.

\begin{figure}[!t]
\includegraphics[width=0.8\linewidth]{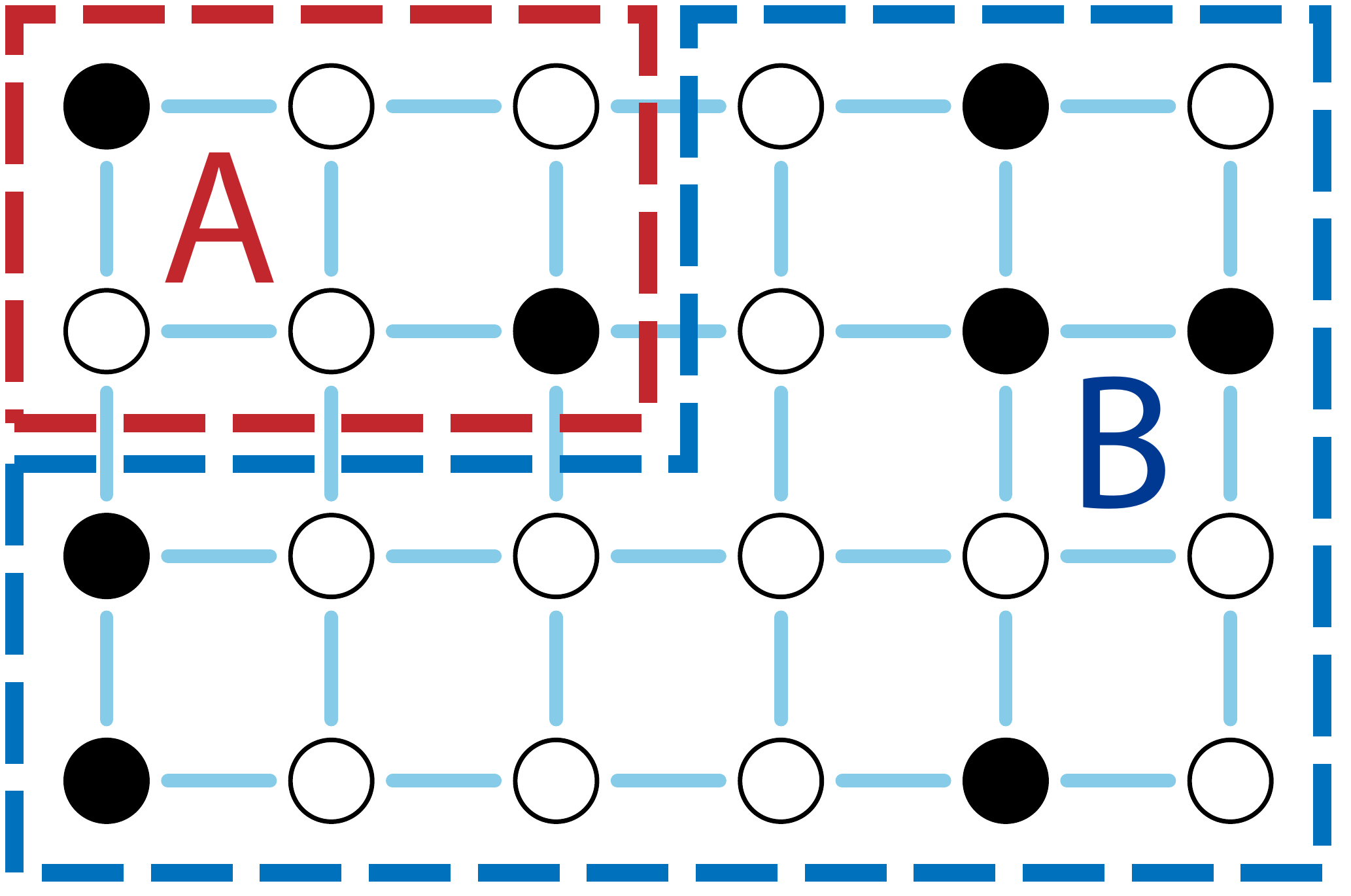}\\
\caption{A 2-dimensional lattice of size $L=24$ sites and $n=8$ particles is shown. The subsystems $A$ and $B$ are also depicted as red and blue regions respectively. The smaller subsystem $A$ has $M=6$ sites and $n_A=2$ particles in this example.}
\label{fig:lattice2D}
\end{figure}

\begin{figure}[!t]
\includegraphics[width=0.8\linewidth]{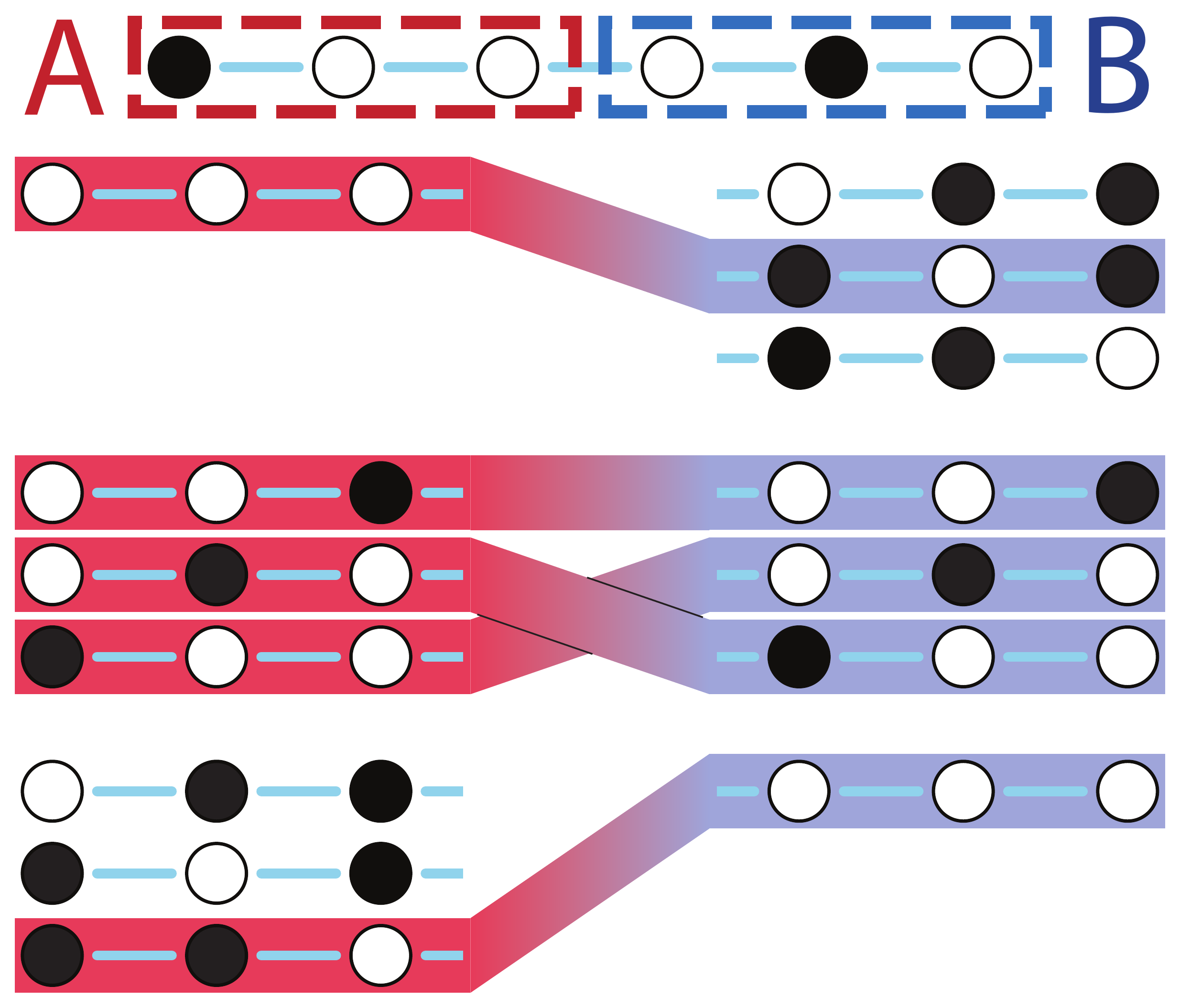}\\
\caption{A maximally entangled state is such where one has the maximal uncertainty about the state of the full system, but determining the state of subsystem $A$ also determines the state of subsystem $B$ with certainty. This means that when constructing such a state, none of the orthogonal states spanning subsystems $A$ and $B$ can be used twice. But since the conservation law prohibits matching states whose particle numbers do not add up to the total number of particles, the maximal entanglement entropy is lower than initially expected. In this figure, one of the maximally entangled states $\ket{\psi}=\frac{1}{\sqrt{5}}(\ket{000101}+\ket{001001}+\ket{010100}+\ket{100010}+\ket{110000})$ for the example mentioned in the introduction is shown, leading to $S_{\mathrm{ent}}^{(\max)}=\ln\big(\binom{3}{0}+\binom{3}{1}+\binom{3}{0}\big)=\ln 5$.}
\label{fig:maxentstate}
\end{figure}

\emph{Bound on entanglement entropy.---}
For a bipartite system of $n$ spinless particles moving on a system of $L$ number of sites, which is partitioned into two subsystems $A$ and $B$ ($\HS=\HS_A\otimes \HS_B$) with $M$ and $L-M$ number of sites (see Fig.~\ref{fig:lattice2D}), assuming that the state of the composite system is pure and that $n\leq M\leq L-M$, the entanglement entropy is bounded by
\[
S_{\mathrm{ent}}\leq \ln \sum_{\na=0}^n \min\Big\{\dim \HS_A^{(\na)},\dim \HS_B^{(n-\na)}\Big\},
\]
where  $\HS_A^{(\na)}$ denotes the Hilbert spaces of exactly $\na$ particles contained in the subsystem $A$, and $\HS_B^{(n-\na)}$ denotes the Hilbert space of exactly $n-\na$ particles contained in subsystem $B$ respectively. This is a tight bound, meaning that it can be saturated with a specific wave function of $n$ particles. An application of this result is shown in Fig.~\ref{fig:maxentstate}.

The above formula can be generalized to include cases $n>M$, but the fermionic and bosonic cases must be treated separately. For fermionic systems (or systems of hard-core bosons) $\dim \HS_A^{(\na)}=\binom{M}{\na}$ which leads to
\[\label{eq:fermionicbound}
S_{\mathrm{ent}}\leq \ln \sum_{\na=\max\{0,n-L+M\}}^{\min\{n,M\}} \min\bigg\{\binom{M}{\na},\binom{L-M}{n-\na}\bigg\},
\]
while for bosonic systems $\dim \HS_A^{(\na)}=\binom{M+\na-1}{\na}$ which leads to
\[\label{eq:bosonicbound}
S_{\mathrm{ent}}\leq \ln \sum_{\na=0}^n \min\bigg\{\binom{M+\na-1}{\na},\binom{L-M+n-\na-1}{n-\na}\bigg\}.
\]
\begin{proof}
Assuming that $n\leq M$, the Hilbert space of $n$ particles contained on lattice of $L$ sites can be decomposed as
\[\label{eq:originaldecomposition}
\HS=\bigoplus_{\na=0}^n\HS_A^{(\na)}\otimes \HS_B^{(n-\na)}.
\]
This means that any wavefunction $\ket{\psi}\in\HS$ can be written as
\[
\ket{\psi}=\sum_{\na=0}^n a_{\na}\ket{\psi_\na},
\]
where $\ket{\psi_\na}\in \HS_A^{(\na)}\otimes \HS_B^{(n-\na)}$. Applying the Schmidt decomposition, we can write each of these vectors as
\[\label{eq:decomposition}
\ket{\psi_\na}=\sum_{i=1}^{d_{\na}} b_i^{(\na)}\ket{\chi_i^{(\na)}}\otimes\ket{\phi_i^{(n-\na)}}
\]
where $d_\na=\min\big\{\dim \HS_A^{(\na)}, \dim \HS_B^{(n-\na)}\big\}$, and $\{\ket{\chi_i^{(\na)}}\}_{i=1}^{d_{\na}}$ and $\{\ket{\phi_i^{(n-\na)}}\}_{i=1}^{d_{\na}}$ form orthogonal sets%.
, and $\{b_i^{(\na)}\}_{i=1}^{d_{\na}}$ are real, non-negative scalars. Also any two vectors $\ket{\chi_i^{({\na})}}$ and $\ket{\chi_j^{(\tilde{n}_{\!A})}}$, $\na\neq \tilde{n}_{\!A}$, are orthogonal to each other, because they belong into subspaces associated with different eigenvalues $\na$ of a Hermitian operator $\hat{N}_A$ (measuring the number of particles in sublattice $A$). The same argument can be made for vectors $\ket{\phi_i^{(n-\na)}}$ using $\hat{N}_B$. This allows us to compute the reduced density matrix,
\[
\R_A=\tr_B[\pro{\psi}{\psi}]=\sum_{\na=0}^n\sum_{i=1}^{d_{\na}}\abs{a_{\na}}^2\abs{b_i^{(\na)}}^2\pro{\chi_i^{(\na)}}{\chi_i^{(\na)}},
\]
and since vectors $\ket{\chi_i^{(\na)}}$ are orthogonal to each other, we can also compute the entanglement entropy as
\[
S_{\mathrm{ent}}\equiv S(\R_A)=-\sum_{\na=0}^n\sum_{i=1}^{d_{\na}}\abs{a_{\na}}^2\abs{b_i^{(\na)}}^2 \ln \abs{a_{\na}}^2\abs{b_i^{(\na)}}^2.
\]
Using Jensen's theorem on the strictly concave function $f(x)=\ln x$, which is a standard procedure for bounding the Shannon entropy, we derive
\[
S_{\mathrm{ent}}\leq \ln \sum_{\na=0}^n d_\na,
\]
which proves the theorem for $n\leq M$. The inequality is saturated if and only if all the probabilities are equal, i.e.,
\[\label{eq:state_of_maximum_EE}
\abs{a_{\na}}^2\abs{b_i^{(\na)}}^2=\bigg(\!\sum_{\na=0}^n d_\na\bigg)^{-1}
\]
for all $\na$ and $i$. Considering decomposition~\eqref{eq:decomposition}, this equation is the sufficient and necessary condition for the state to be maximally entangled in a closed system.

Now let us include cases of $n\geq M$. For a fermionic system, the three cases to consider are: $n\leq M\leq L-M$, $M\leq n \leq L-M$, and $M\leq L-M < n$. Combined, for any $n\leq L$ the Hilbert space can be decomposed as
\[
\HS=\bigoplus_{\na=\max\{0,n-L+M\}}^{\min\{n,M\}} \HS_A^{(\na)}\otimes \HS_B^{(n-\na)}.
\]
The rest of the analysis proceeds analogously and leads to
\[\label{eq:fermionicboundunfiltered}
S_{\mathrm{ent}}\leq \ln\!\!\! \sum_{\na=\max\{0,n-L+M\}}^{\min\{n,M\}}\!\!\! \min\Big\{\dim \HS_A^{(\na)},\dim \HS_B^{(n-\na)}\Big\},
\]
with equality if and only if $\abs{a_{\na}}^2\abs{b_i^{(\na)}}^2=\big(\!\sum_{\na=\max\{0,n-L+M\}}^{\min\{n,M\}} d_\na\big)^{-1}$ for all $\na$ and $i$. Considering that $\dim \HS_A^{(\na)}=\binom{M}{\na}$ (combination: the number of ways we can distribute $n_A$ particles in a sublattice of $M$ sites, where no repetition is possible due to the Pauli exclusion principle or hard-core condition) and $\dim \HS_B^{(n-\na)}=\binom{L-M}{n-\na}$, we obtain Eq.~\eqref{eq:fermionicbound}.

For a bosonic system, the decomposition of Hilbert space is identical to Eq.~\eqref{eq:originaldecomposition} for any $n$. The formula therefore remains the same, and considering that for a bosonic system we have $\dim \HS_A^{(\na)}=\binom{M+\na-1}{\na}$ (combination with repetition: the number of ways we can distribute $n_A$ particles in a sublattice of $M$ sites, where multiple particles can be in a single site) and $\dim \HS_B^{(n-\na)}=\binom{L-M+n-\na-1}{n-\na}$, we obtain Eq.~\eqref{eq:bosonicbound}.
\end{proof}

The condition for the maximally entangled state, Eq.~\eqref{eq:state_of_maximum_EE}, has an interesting implication. It gives prediction for the number of particles in each of the subsystems: if the state is maximally entangled, then the probability of measuring $\na$ particles in sublattice $A$ (which must be the same as the probability of measuring $n-\na$ particles in sublattice $B$) is equal to
\[\label{eq:fermprob}
p_{\na}=\abs{a_{\na}}^2=\frac{d_\na}{\sum_{\na=\max\{0,n-L+M\}}^{\min\{n,M\}}  d_\na},
\] 
$d_\na=\min\big\{\binom{M}{\na},\binom{L-M}{n-\na}\big\}$, for the fermionic gas, and
\[\label{eq:bosprob}
p_{\na}=\abs{a_{\na}}^2=\frac{d_\na}{\sum_{\na=0}^n  d_\na},
\]
$d_\na=\min\big\{\binom{M+\na-1}{\na},\binom{L-M+n-\na-1}{n-\na}\big\}$, for the bosonic gas. The mean number of particles in sublattice $A$ is $\ov{\na}=\sum_{\na=\max\{0,n-L+M\}}^{\min\{n,M\}} p_{\na}\na$ and $\ov{\na}=\sum_{\na=0}^n p_{\na}\na$ (while $\ov{n_{\!B}}=n-\ov{\na}$) for the fermionic and the bosonic gas respectively. Therefore, if a state of a closed system does not satisfy these properties, it cannot be maximally entangled~\footnote{This is necessary, but not a sufficient condition for the state to be maximally entangled. This means that even if the Eqs.~\eqref{eq:fermprob} or \eqref{eq:bosprob} are satisfied, the state does not have to be maximally entangled. The sufficient and necessary condition is given by Eq.~\eqref{eq:state_of_maximum_EE} for $n\leq M$ and its generalizations for cases $n\geq M$}.

One can also notice that the derived bound stops depending on the total system size $L$ if it is large enough. Specifically, for fermionic systems and
\[\label{eq:conditionforL}
L\geq \max\bigg\{\max_{\na\in\{0,\dots,\min\{n,M\}\}}\bigg\{\binom{M}{\na}\bigg\},n\bigg\}+M,
\]
the bound becomes
\[\label{eq:flattening}
S_{\mathrm{ent}}\leq \ln\bigg(1+ \sum_{\na=0}^{\min\{n,M\}-1} \binom{M}{\na}\bigg),
\]
which no longer depends on $L$.

If in addition $n\geq M$, then
\[
S_{\mathrm{ent}}\leq \ln\sum_{\na=0}^{M} \binom{M}{\na}=\ln 2^M.
\]
which is equal to the maximal entropy of subsystem $\HS_A$. This is the same result that could be recovered from the original bound, Eq.~\eqref{eq:originalbound}. Therefore, for fermionic systems with large enough baths (subsystems $B$), and a large number of particles, these bounds are the same. The same \emph{does not} hold for bosonic systems however, for which $S_{\mathrm{ent}}\leq \ln\Big(1+ \sum_{\na=0}^{n-1} \binom{M+\na-1}{\na}\Big)<\ln \sum_{\na=0}^{n} \binom{M+\na-1}{\na}=\ln \dim \HS_A$, irrespective of $n$, for large $L$ and $M>1$. Thus for closed bosonic systems, our bound is always better.

It also turns out that Eq.~\eqref{eq:flattening} is the value of the bound in the thermodynamic limit, where both the number of particles $n$ and size of the system $L$ grow to infinity, but the particle density $c=n/L$ remains constant, while keeping $M$ constant. This can be shown by dividing condition~\eqref{eq:conditionforL} by $n$ and taking the limit, which gives $c\leq 1$, which must be by definition satisfied for any spinless fermionic system.

\begin{figure}[!t]
\includegraphics[width=0.9\linewidth]{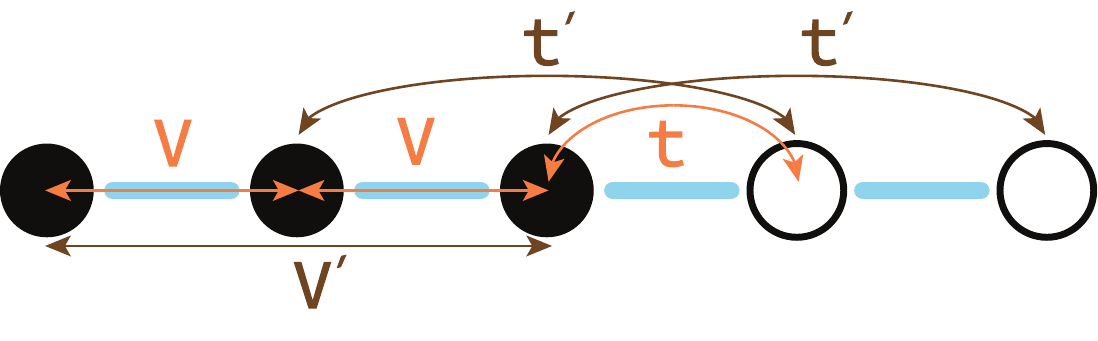}
\caption{A 1-dimensional lattice of size $L=5$ sites and $n=3$ particles is shown. The right hand side of the figure illustrates the hopping terms $t$ or $t'$ i.e., particles move to the nearest-neighbor (NN) and next-nearest-neighbor (NNN) sites respectively. The left hand side of the figure shows the interactions of strengths $V$ and $V'$ between NN and NNN respectively.}
\label{fig:lattice1D}
\end{figure}

\emph{Achievability of the bound in 1D fermionic lattice.---}
Here, we illustrate the derived upper bound~\eqref{eq:fermionicbound} in a simulation. We specifically focus on the case where $n<M\leq L-M$. The other cases turned out to be very similar, and we shall not show them here.

We consider a system of $n$ spin-less fermions in a 1-dimensional lattice of size $L$, with Hamiltonian
\[
\begin{split}
\hat{H} = \sum_{i=1}^{L} [-t({f}_{i}^{\dagger}{f}_{i+1}+h.c.)+V{n}_{i}^{f}{n}_{i+1}^{f}\\
-t'({f}_{i}^{\dagger}{f}_{i+2}+h.c.)+V'{n}_{i}^{f}{n}_{i+2}^{f}],
\end{split}
\]
where $f_{i}$ and $f_{i}^{\dagger}$ are fermionic annihilation and creation operators for site $i$ and ${n}_{i}^{f} = {f}_{i}^{\dagger}{f}_{i}$ is the local density operator. The nearest-neighbor (NN) and next-nearest-neighbor (NNN) hopping terms are respectively $t$ and $t'$ and the interaction strengths are $V$ and $V'$ as illustrated in Fig.~\ref{fig:lattice1D}. We choose this Hamiltonian since it has been extensively studied in the literature~
\cite{rigol2008thermalization,santos2010onset,santos2012weak,deutsch2013microscopic,beugeling2014finite,alba2015eigenstate}, and because it is an archetypal example of both non-integrable (generic; $t',V'\neq 0$) and integrable ($t'=V'=0$) quantum systems.

\begin{figure}[t]
\includegraphics[width=1\linewidth]{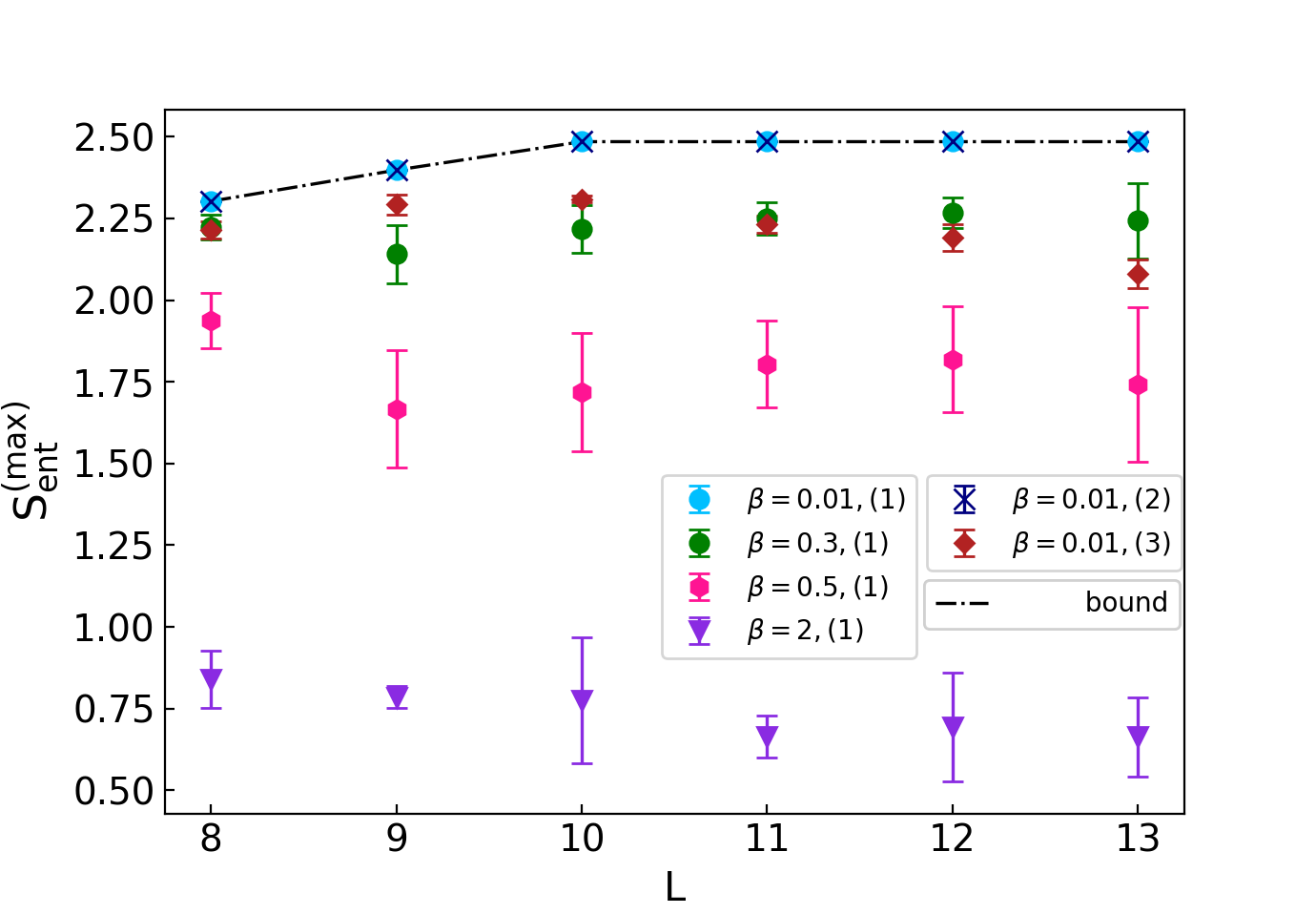}\\
\caption{$S_{\mathrm{ent}}^{(\max)}=\max_{\tau}S_{\mathrm{ent}}(\ket{\psi_{\tau}})$ for different values of $L$ and  $\beta$, $n=3$ particles, for a subsystem $A$ being fixed as the left $M=4$ sites of the chain. In the low temperature limit, $\beta=2$, the initial state is close to being a ground state. In the high temperature limit, $\beta=0.01$, the initial state becomes a \emph{random pure state}~\cite{lubkin1978entropy,lloyd1988complexity,page1993average,dahlsten2014entanglement},
in which all the energy coefficients are equal on average. We show cases of (1) non-integrable system  ($t=t'=1.9$, $V=V'=0.5$) and varying temperatures, and cases of high temperature with (2) NN hopping only ($t=1.9$, $t'=V=V'=0$), and with (3) interaction only ($V=V'=0.5$, $t=t'=0$). In cases (1) and (2) and high temperature, the maximum value of $S_{\mathrm{ent}}$ reaches exactly the theoretical bound (dashed line) for all $L$, but not in case (3). This shows that the non-zero hopping term is the most important for reaching the maximum.}
\label{fig:maxSent}
\end{figure}

In the simulation depicted in Fig.~\ref{fig:maxSent}, we take $t=t'=1.9$, $V=V'=0.5$~\footnote{It does not matter much which particular values we choose, as long as $t,t',V,V'\neq 0$, the evolution is qualitatively the same.}, and cases with NN hopping only, and with interaction only. The total number of particles is $n=3$, and we take subsystem $A$ to be the $M=4$ sites on the left side of the chain, while the full system size $L$, and inverse temperature $\beta=1/T$ are both varied.

We take the initial state to be the complex \emph{random pure thermal state} (RPTS) (also known as the thermal pure quantum or canonical thermal pure quantum state~\cite{sugiura2012thermal,sugiura2013canonical,nakagawa2018universality}), which we define as
\[
\ket{\psi}=\frac{1}{\sqrt{Z}}\sum_E c_E e^{-\beta E/2}\ket{E},
\]
where $\ket{E}$'s are the eigenstates of the total Hamiltonian, computed using exact diagonalization. The coefficients $\{c_E\}$ are random complex numbers, $c_E\equiv (x_E+iy_E)/\sqrt{2}$, with $x_E$ and $y_E$ obeying the standard normal distribution $\mathcal{N}(0,1)$, and $Z=\sum_E \abs{c_E}^2 e^{-\beta E}$ is the normalization constant. This state emulates a thermal state, while being pure. This initial state is entangled but not maximally entangled. For instance, in the case of $\beta=0.01$ and a given initial RPTS, the initial entanglement entropies associated with system sizes $L=[8,9,10,11,12,13]$ are $S_{\mathrm{ent}}=[2.031,2.091,2.118,2.122,2.080,2.020]$. The initial state is then evolved as $\ket{\psi_{\tau}}=e^{-i \hat{H}\tau}\ket{\psi}$.

To find the maximum value that $S_{\ent}$ can achieve during its time evolution, we use the simplex search algorithm. For a given $L$ and $\beta$, we initialize the state in $6$ different complex RPTS, and find the maxima for each initial state by maximizing over phases $\phi_E=E \tau$ that appear in the time evolution of the wavefunction, $\hat{U}_{\tau} = e^{-i\hat{H}\tau}$. As long as the differences of  $E$'s are irrational (or close to being to), this method must give the same result as maximizing over all times $\tau$. We do this because maximizing over the time by simply evolving the system would take an extremely long time. We then plot the mean value of these six maxima as well as the standard deviation (depicted as error bars) in Fig.~\ref{fig:maxSent}. The theoretical bound, Eq.~\eqref{eq:fermionicbound} for the case where $n<M\leq L-M$, is plotted in the same figure (dashed line) for various values of $L$.

As Fig.~\ref{fig:maxSent} illustrates, in the non-integrable case for $\beta=0.01$ the theoretical bound~\eqref{eq:fermionicbound} is saturated exactly. For large $L$ the bound flattens, as expected from Eq.~\eqref{eq:flattening}.
As the temperature drops, the system can no longer achieve this bound, but the maximum entanglement entropy still stays approximately constant for large $L$. Interestingly, we found that the upper bound is reached during the unitary evolution also for integrable systems with NN interaction and hopping ($t'=V'=0$), or even for systems with just the NN hopping term ($t'=V=V'=0$), but not when there is no hopping, which we can summarize as ``As long as there is some hopping, in both cases of integrable or non-integrable systems, the bound is achieved during the unitary evolution, if the temperature is high enough.''

Regarding the average number of particles in the subsystem, states for which $S_{\mathrm{ent}}$ has reached its maximum were measured to have $\ov{\na}= [1.50,1.55,1.58,1.58,1.58,1.58]$ for $L=[8,9,10,11,12,13]$, which corresponds exactly to the prediction of Eq.~\eqref{eq:fermprob}.

\emph{Conclusion and applications.---}
We derived a tight upper bound on entanglement entropy for bipartite systems with a conserved number of spinless particles. We showed numerically that at high temperature, the maximum entanglement entropy of a fermionic lattice in fact reaches this upper bound during its time evolution. Furthermore, by studying the maximally entangled states, we found that measuring the particle number in one of the subsystems can serve as a simple test of whether the state can be maximally entangled. 
In contrast to Ref.~\cite{vidmar2017entanglementquadratic}, which derived bounds for the average entanglement entropy of all eigenstates of quadratic fermionic Hamiltonians, our bound holds for any state in any spinless fermionic and bosonic system with a conserved number of particles, irrespective of the Hamiltonian being used.

Our results can be also directly transferred to lattices of identical spin-1/2 particles with the total spin conserved, where spin-up and spin-down take the place of a particle and a hole respectively, or to lattices of qubits consisting of different energy states (such as cold atoms~\cite{bloch2008many,islam2015measuring,kaufman2016quantum,browaeys2016experimental}, trapped ions~\cite{zhang2017observation,friis2018observation,brydges2018probing}, or superconducting qubits~\cite{fitzpatrick2017observation,gambetta2017building,xu2018emulating,neill2018blueprint}), when the total energy, and therefore also the total number of excited states is conserved, while neglecting the interaction energy.

Using that for a pure state $\R_{AB}$, $S(\R_{AB})=0$ and $S(\R_A)=S(\R_B)$, a noteworthy consequence of the bound, Eqs.~\eqref{eq:fermionicbound} and~\eqref{eq:bosonicbound}, is that it sets a lower bound on conditional entropy $S(A|B)_{\R} = S(\R_{AB}) - S(\R_B)$~\cite{cerf1997negative,preskill2016quantum,winter2016tight} which gives a sufficient and necessary condition for teleportation~\cite{witten2018mini}, and an upper bound on the mutual information $I(A;B)=S(\R_A)+S(\R_B)-S(\R_{AB})$~\cite{um2012entanglement,preskill2016quantum,de2017quantum}, which determines the largest possible rate of communication~\cite{ohya2010quantum,dixon2012quantum,wilde2017}, and has applications in quantum machine learning~\cite{turkpencce2017quantum,datta2018convexity,koch2018mutual}.

Another implication of this result is with regards to R\'enyi entropies of higher order. For a general density matrix, entanglement entropy (R\'enyi entropy of order $\alpha=1$) is related to R\'enyi entropies of higher order, $S_{\alpha>1}$, by inequality $S_{\ent}(\R) \geq S_{\alpha>1}(\R)$~\cite{fannes2012connecting}. This means that the upper bound on entanglement entropy found in this Rapid Communication could be taken as a loose upper bound on $S_{\alpha>1}(\R)$.

This is important due to the existence of experiments involving measurements on R\'enyi-2 entanglement entropy $S_{\alpha=2}$~\cite{islam2015measuring,kaufman2016quantum,brydges2018probing}, which allows us to compare our bound with experimental data. Ref.~\cite{islam2015measuring} used a system of ultracold bosonic atoms trapped in an optical lattice, evolving by the Bose–-Hubbard Hamiltonian. The maximum R\'enyi entropy of a ground state for a system of $L=4$ sites and $n=4$ particles, and various sizes of subsystems $M=[1,2,4]$ was obtained from Fig.~4 in~\cite{islam2015measuring} (including an offset of about $0.5$) as $S_{\alpha=2}=[0.6,0.9,0]$, which is below the bound $S_{\ent}^{(\mathrm{bound})}=[1.6,2.2,0]$ calculated from Eq.~\eqref{eq:bosonicbound}. The maximal achieved entropy obtained from Fig.~6 in~\cite{islam2015measuring} for $L=n=2$ and $M=1$ is $S_{\alpha=2}=0.8$ which is much closer to the bound $S_{\ent}^{(\mathrm{bound})}=1.1$. Ref.~\cite{kaufman2016quantum} focused on measuring the R\'enyi entropy of an evolving system using the same model. The maximum values of the R\'enyi entropy read out from Fig.~3 in~\cite{kaufman2016quantum} for $L=n=6$ and $M=[1,2,3,6]$ are $S_{\alpha=2}=[0.8,1.9,2.6,0]$, while the bound gives $S_{\ent}^{(\mathrm{bound})}=[1.9,3.0,3.4,0]$. Finally, Ref.~\cite{brydges2018probing} used a system of trapped ions, each carrying a spin, evolved by an XY Hamiltonian which conserves the total spin. This model is therefore mathematically identical to a lattice of spinless fermions. $L=10$ atoms were prepared in the N\'eel state ($n=5$), and after $5\ \mathrm{ms}$ the R\'enyi entropy was read out for $M=[1,2,3,4,5,6,7,8,9]$ (Fig.~2 in~\cite{brydges2018probing}) at values scattered around $S_{\alpha=2}\approx[0.6,1.3,1.7,2.1,2.4,2.3,1.9,1.5,0.8]$ (recalculated by changing the base of logarithm $\log_2\rightarrow \ln$). These values are comparable but two of them are slightly higher than the bound $S_{\ent}^{(\mathrm{bound})}=[0.7,1.4,2.1,2.8,3.5,2.8,2.1,1.4,0.7]$ calculated from Eq.~\eqref{eq:fermionicbound}, due to inadvertently introduced decoherence (the total R\'enyi entropy was $0.5$ at the time of measurement).

We conclude that our results are in accordance with current experiments, and will become especially useful for bounding the amount of entanglement spontaneously created in closed systems, once a direct measurement of entanglement entropy becomes feasible.

\acknowledgements{We are grateful to Joshua M. Deutsch and Anthony Aguirre for fruitful discussions, and thank Joshua M. Deutsch for providing his code for simulating 1-dimensional fermionic lattice. D\v{S} acknowledges support from Foundational Questions Institute (FQXi.org) and from the Faggin Presidential Chair Fund.}

\bibliographystyle{apsrev4-1}
\bibliography{bibl1}

\begin{thebibliography}{103}%
\makeatletter
\providecommand \@ifxundefined [1]{%
 \@ifx{#1\undefined}
}%
\providecommand \@ifnum [1]{%
 \ifnum #1\expandafter \@firstoftwo
 \else \expandafter \@secondoftwo
 \fi
}%
\providecommand \@ifx [1]{%
 \ifx #1\expandafter \@firstoftwo
 \else \expandafter \@secondoftwo
 \fi
}%
\providecommand \natexlab [1]{#1}%
\providecommand \enquote  [1]{``#1''}%
\providecommand \bibnamefont  [1]{#1}%
\providecommand \bibfnamefont [1]{#1}%
\providecommand \citenamefont [1]{#1}%
\providecommand \href@noop [0]{\@secondoftwo}%
\providecommand \href [0]{\begingroup \@sanitize@url \@href}%
\providecommand \@href[1]{\@@startlink{#1}\@@href}%
\providecommand \@@href[1]{\endgroup#1\@@endlink}%
\providecommand \@sanitize@url [0]{\catcode `\\12\catcode `\$12\catcode
  `\&12\catcode `\#12\catcode `\^12\catcode `\_12\catcode `\%12\relax}%
\providecommand \@@startlink[1]{}%
\providecommand \@@endlink[0]{}%
\providecommand \url  [0]{\begingroup\@sanitize@url \@url }%
\providecommand \@url [1]{\endgroup\@href {#1}{\urlprefix }}%
\providecommand \urlprefix  [0]{URL }%
\providecommand \Eprint [0]{\href }%
\providecommand \doibase [0]{http://dx.doi.org/}%
\providecommand \selectlanguage [0]{\@gobble}%
\providecommand \bibinfo  [0]{\@secondoftwo}%
\providecommand \bibfield  [0]{\@secondoftwo}%
\providecommand \translation [1]{[#1]}%
\providecommand \BibitemOpen [0]{}%
\providecommand \bibitemStop [0]{}%
\providecommand \bibitemNoStop [0]{.\EOS\space}%
\providecommand \EOS [0]{\spacefactor3000\relax}%
\providecommand \BibitemShut  [1]{\csname bibitem#1\endcsname}%
\let\auto@bib@innerbib\@empty
%</preamble>
\bibitem [{\citenamefont {Nikoli{\'c}}(2012)}]{nikolic2012epr}%
  \BibitemOpen
  \bibfield  {author} {\bibinfo {author} {\bibfnamefont {H.}~\bibnamefont
  {Nikoli{\'c}}},\ }\href@noop {} {\bibfield  {journal} {\bibinfo  {journal}
  {\href{https://iopscience.iop.org/article/10.1088/0143-0807/33/5/1089/meta}{Eur.
  J. Phys.}}\ }\textbf {\bibinfo {volume} {33}},\ \bibinfo {pages} {1089}
  (\bibinfo {year} {2012})}\BibitemShut {NoStop}%
\bibitem [{\citenamefont {Osborne}\ and\ \citenamefont
  {Nielsen}(2002)}]{osborne2002entanglement}%
  \BibitemOpen
  \bibfield  {author} {\bibinfo {author} {\bibfnamefont {T.~J.}\ \bibnamefont
  {Osborne}}\ and\ \bibinfo {author} {\bibfnamefont {M.~A.}\ \bibnamefont
  {Nielsen}},\ }\href@noop {} {\bibfield  {journal} {\bibinfo  {journal}
  {\href{https://journals.aps.org/pra/abstract/10.1103/PhysRevA.66.032110}{Phys.
  Rev. A}}\ }\textbf {\bibinfo {volume} {66}},\ \bibinfo {pages} {032110}
  (\bibinfo {year} {2002})}\BibitemShut {NoStop}%
\bibitem [{\citenamefont {Amico}\ \emph {et~al.}(2008)\citenamefont {Amico},
  \citenamefont {Fazio}, \citenamefont {Osterloh},\ and\ \citenamefont
  {Vedral}}]{amico2008entanglement}%
  \BibitemOpen
  \bibfield  {author} {\bibinfo {author} {\bibfnamefont {L.}~\bibnamefont
  {Amico}}, \bibinfo {author} {\bibfnamefont {R.}~\bibnamefont {Fazio}},
  \bibinfo {author} {\bibfnamefont {A.}~\bibnamefont {Osterloh}}, \ and\
  \bibinfo {author} {\bibfnamefont {V.}~\bibnamefont {Vedral}},\ }\href@noop {}
  {\bibfield  {journal} {\bibinfo  {journal}
  {\href{https://journals.aps.org/rmp/abstract/10.1103/RevModPhys.80.517}{ Rev.
  Mod. Phys.,}}\ }\textbf {\bibinfo {volume} {80}},\ \bibinfo {pages} {517}
  (\bibinfo {year} {2008})}\BibitemShut {NoStop}%
\bibitem [{\citenamefont {Bloch}\ \emph {et~al.}(2008)\citenamefont {Bloch},
  \citenamefont {Dalibard},\ and\ \citenamefont {Zwerger}}]{bloch2008many}%
  \BibitemOpen
  \bibfield  {author} {\bibinfo {author} {\bibfnamefont {I.}~\bibnamefont
  {Bloch}}, \bibinfo {author} {\bibfnamefont {J.}~\bibnamefont {Dalibard}}, \
  and\ \bibinfo {author} {\bibfnamefont {W.}~\bibnamefont {Zwerger}},\
  }\href@noop {} {\bibfield  {journal} {\bibinfo  {journal}
  {\href{https://journals.aps.org/rmp/abstract/10.1103/RevModPhys.80.885}{Rev.
  Mod. Phys.}}\ }\textbf {\bibinfo {volume} {80}},\ \bibinfo {pages} {885}
  (\bibinfo {year} {2008})}\BibitemShut {NoStop}%
\bibitem [{\citenamefont {Lee}(2009)}]{lee2009lattice}%
  \BibitemOpen
  \bibfield  {author} {\bibinfo {author} {\bibfnamefont {D.}~\bibnamefont
  {Lee}},\ }\href@noop {} {\bibfield  {journal} {\bibinfo  {journal}
  {\href{https://www.sciencedirect.com/science/article/pii/S014664100800094X}{Prog.
  Part. Nucl. Phys.}}\ }\textbf {\bibinfo {volume} {63}},\ \bibinfo {pages}
  {117} (\bibinfo {year} {2009})}\BibitemShut {NoStop}%
\bibitem [{\citenamefont {Zhang}\ \emph {et~al.}(2011)\citenamefont {Zhang},
  \citenamefont {Grover},\ and\ \citenamefont
  {Vishwanath}}]{zhang2011entanglement}%
  \BibitemOpen
  \bibfield  {author} {\bibinfo {author} {\bibfnamefont {Y.}~\bibnamefont
  {Zhang}}, \bibinfo {author} {\bibfnamefont {T.}~\bibnamefont {Grover}}, \
  and\ \bibinfo {author} {\bibfnamefont {A.}~\bibnamefont {Vishwanath}},\
  }\href@noop {} {\bibfield  {journal} {\bibinfo  {journal}
  {\href{https://journals.aps.org/prl/abstract/10.1103/PhysRevLett.107.067202}{Phys.
  Rev. Lett.}}\ }\textbf {\bibinfo {volume} {107}},\ \bibinfo {pages} {067202}
  (\bibinfo {year} {2011})}\BibitemShut {NoStop}%
\bibitem [{\citenamefont {Jiang}\ \emph {et~al.}(2012)\citenamefont {Jiang},
  \citenamefont {Wang},\ and\ \citenamefont {Balents}}]{jiang2012identifying}%
  \BibitemOpen
  \bibfield  {author} {\bibinfo {author} {\bibfnamefont {H.-C.}\ \bibnamefont
  {Jiang}}, \bibinfo {author} {\bibfnamefont {Z.}~\bibnamefont {Wang}}, \ and\
  \bibinfo {author} {\bibfnamefont {L.}~\bibnamefont {Balents}},\ }\href@noop
  {} {\bibfield  {journal} {\bibinfo  {journal}
  {\href{https://www.nature.com/articles/nphys2465}{Nat. Phys.}}\ }\textbf
  {\bibinfo {volume} {8}},\ \bibinfo {pages} {902} (\bibinfo {year}
  {2012})}\BibitemShut {NoStop}%
\bibitem [{\citenamefont {Ekert}(1991)}]{ekert1991quantum}%
  \BibitemOpen
  \bibfield  {author} {\bibinfo {author} {\bibfnamefont {A.~K.}\ \bibnamefont
  {Ekert}},\ }\href@noop {} {\bibfield  {journal} {\bibinfo  {journal}
  {\href{https://journals.aps.org/prl/abstract/10.1103/PhysRevLett.67.661}{Phys.
  Rev. Lett.}}\ }\textbf {\bibinfo {volume} {67}},\ \bibinfo {pages} {661}
  (\bibinfo {year} {1991})}\BibitemShut {NoStop}%
\bibitem [{\citenamefont {Bennett}\ \emph {et~al.}(1993)\citenamefont
  {Bennett}, \citenamefont {Brassard}, \citenamefont {Cr{\'e}peau},
  \citenamefont {Jozsa}, \citenamefont {Peres},\ and\ \citenamefont
  {Wootters}}]{bennett1993teleporting}%
  \BibitemOpen
  \bibfield  {author} {\bibinfo {author} {\bibfnamefont {C.~H.}\ \bibnamefont
  {Bennett}}, \bibinfo {author} {\bibfnamefont {G.}~\bibnamefont {Brassard}},
  \bibinfo {author} {\bibfnamefont {C.}~\bibnamefont {Cr{\'e}peau}}, \bibinfo
  {author} {\bibfnamefont {R.}~\bibnamefont {Jozsa}}, \bibinfo {author}
  {\bibfnamefont {A.}~\bibnamefont {Peres}}, \ and\ \bibinfo {author}
  {\bibfnamefont {W.~K.}\ \bibnamefont {Wootters}},\ }\href@noop {} {\bibfield
  {journal} {\bibinfo  {journal}
  {\href{https://journals.aps.org/prl/abstract/10.1103/PhysRevLett.70.1895}{Phys.
  Rev. Lett.}}\ }\textbf {\bibinfo {volume} {70}},\ \bibinfo {pages} {1895}
  (\bibinfo {year} {1993})}\BibitemShut {NoStop}%
\bibitem [{\citenamefont {Nielsen}\ and\ \citenamefont
  {Chuang}(2002)}]{nielsen2002quantum}%
  \BibitemOpen
  \bibfield  {author} {\bibinfo {author} {\bibfnamefont {M.~A.}\ \bibnamefont
  {Nielsen}}\ and\ \bibinfo {author} {\bibfnamefont {I.}~\bibnamefont
  {Chuang}},\ }\href@noop {} {\enquote {\bibinfo {title} {Quantum computation
  and quantum information},}\ } (\bibinfo {year} {2002})\BibitemShut {NoStop}%
\bibitem [{\citenamefont {Wendin}(2017)}]{wendin2017quantum}%
  \BibitemOpen
  \bibfield  {author} {\bibinfo {author} {\bibfnamefont {G.}~\bibnamefont
  {Wendin}},\ }\href@noop {} {\bibfield  {journal} {\bibinfo  {journal}
  {\href{https://iopscience.iop.org/article/10.1088/1361-6633/aa7e1a/meta}{Rep.
  Prog. Phys.}}\ }\textbf {\bibinfo {volume} {80}},\ \bibinfo {pages} {106001}
  (\bibinfo {year} {2017})}\BibitemShut {NoStop}%
\bibitem [{\citenamefont {Ng}\ and\ \citenamefont
  {Woods}(2018)}]{ng2018resource}%
  \BibitemOpen
  \bibfield  {author} {\bibinfo {author} {\bibfnamefont {N.}~\bibnamefont
  {Ng}}\ and\ \bibinfo {author} {\bibfnamefont {M.~P.}\ \bibnamefont {Woods}},\
  }\enquote {\bibinfo {title} {\href{https://arxiv.org/abs/1805.09564}{Resource
  theory of quantum thermodynamics: Thermal operations and Second Laws}},}\ in\
  \href@noop {} {\emph {\bibinfo {booktitle} {Thermodynamics in the Quantum
  Regime}}}\ (\bibinfo  {publisher} {Springer},\ \bibinfo {year} {2018})\ pp.\
  \bibinfo {pages} {625--650}\BibitemShut {NoStop}%
\bibitem [{\citenamefont {Chitambar}\ and\ \citenamefont
  {Gour}(2019)}]{chitambar2019quantum}%
  \BibitemOpen
  \bibfield  {author} {\bibinfo {author} {\bibfnamefont {E.}~\bibnamefont
  {Chitambar}}\ and\ \bibinfo {author} {\bibfnamefont {G.}~\bibnamefont
  {Gour}},\ }\href@noop {} {\bibfield  {journal} {\bibinfo  {journal}
  {\href{https://journals.aps.org/rmp/abstract/10.1103/RevModPhys.91.025001}{Rev.
  Mod. Phys.}}\ }\textbf {\bibinfo {volume} {91}},\ \bibinfo {pages} {025001}
  (\bibinfo {year} {2019})}\BibitemShut {NoStop}%
\bibitem [{\citenamefont {Landsberg}(2019)}]{landsberg2019very}%
  \BibitemOpen
  \bibfield  {author} {\bibinfo {author} {\bibfnamefont {J.~M.}\ \bibnamefont
  {Landsberg}},\ }\enquote {\bibinfo {title}
  {\href{https://arxiv.org/abs/1801.05893}{A very brief introduction to quantum
  computing and quantum information theory for mathematicians}},}\ in\
  \href@noop {} {\emph {\bibinfo {booktitle} {Quantum Physics and Geometry}}}\
  (\bibinfo  {publisher} {Springer},\ \bibinfo {year} {2019})\ pp.\ \bibinfo
  {pages} {5--41}\BibitemShut {NoStop}%
\bibitem [{\citenamefont {Giovannetti}\ \emph {et~al.}(2006)\citenamefont
  {Giovannetti}, \citenamefont {Lloyd},\ and\ \citenamefont
  {Maccone}}]{giovannetti2006quantum}%
  \BibitemOpen
  \bibfield  {author} {\bibinfo {author} {\bibfnamefont {V.}~\bibnamefont
  {Giovannetti}}, \bibinfo {author} {\bibfnamefont {S.}~\bibnamefont {Lloyd}},
  \ and\ \bibinfo {author} {\bibfnamefont {L.}~\bibnamefont {Maccone}},\
  }\href@noop {} {\bibfield  {journal} {\bibinfo  {journal}
  {\href{https://journals.aps.org/prl/abstract/10.1103/PhysRevLett.96.010401}{Phys.
  Rev. Lett.}}\ }\textbf {\bibinfo {volume} {96}},\ \bibinfo {pages} {010401}
  (\bibinfo {year} {2006})}\BibitemShut {NoStop}%
\bibitem [{\citenamefont {Dowling}(2008)}]{dowling2008quantum}%
  \BibitemOpen
  \bibfield  {author} {\bibinfo {author} {\bibfnamefont {J.~P.}\ \bibnamefont
  {Dowling}},\ }\href@noop {} {\bibfield  {journal} {\bibinfo  {journal}
  {\href{https://www.tandfonline.com/doi/full/10.1080/00107510802091298}{Contemp.
  Phys.}}\ }\textbf {\bibinfo {volume} {49}},\ \bibinfo {pages} {125} (\bibinfo
  {year} {2008})}\BibitemShut {NoStop}%
\bibitem [{\citenamefont {Giovannetti}\ \emph {et~al.}(2011)\citenamefont
  {Giovannetti}, \citenamefont {Lloyd},\ and\ \citenamefont
  {Maccone}}]{giovannetti2011advances}%
  \BibitemOpen
  \bibfield  {author} {\bibinfo {author} {\bibfnamefont {V.}~\bibnamefont
  {Giovannetti}}, \bibinfo {author} {\bibfnamefont {S.}~\bibnamefont {Lloyd}},
  \ and\ \bibinfo {author} {\bibfnamefont {L.}~\bibnamefont {Maccone}},\
  }\href@noop {} {\bibfield  {journal} {\bibinfo  {journal}
  {\href{https://www.nature.com/articles/nphoton.2011.35}{Nat. Photonics}}\
  }\textbf {\bibinfo {volume} {5}},\ \bibinfo {pages} {222} (\bibinfo {year}
  {2011})}\BibitemShut {NoStop}%
\bibitem [{\citenamefont {Demkowicz-Dobrza{\'n}ski}\ \emph
  {et~al.}(2012)\citenamefont {Demkowicz-Dobrza{\'n}ski}, \citenamefont
  {Ko{\l}ody{\'n}ski},\ and\ \citenamefont
  {Gu{\c{t}}{\u{a}}}}]{demkowicz2012elusive}%
  \BibitemOpen
  \bibfield  {author} {\bibinfo {author} {\bibfnamefont {R.}~\bibnamefont
  {Demkowicz-Dobrza{\'n}ski}}, \bibinfo {author} {\bibfnamefont
  {J.}~\bibnamefont {Ko{\l}ody{\'n}ski}}, \ and\ \bibinfo {author}
  {\bibfnamefont {M.}~\bibnamefont {Gu{\c{t}}{\u{a}}}},\ }\href@noop {}
  {\bibfield  {journal} {\bibinfo  {journal}
  {\href{https://www.nature.com/articles/ncomms2067}{Nat. Commun.}}\ }\textbf
  {\bibinfo {volume} {3}},\ \bibinfo {pages} {1063} (\bibinfo {year}
  {2012})}\BibitemShut {NoStop}%
\bibitem [{\citenamefont {T{\'o}th}\ and\ \citenamefont
  {Apellaniz}(2014)}]{toth2014quantum}%
  \BibitemOpen
  \bibfield  {author} {\bibinfo {author} {\bibfnamefont {G.}~\bibnamefont
  {T{\'o}th}}\ and\ \bibinfo {author} {\bibfnamefont {I.}~\bibnamefont
  {Apellaniz}},\ }\href@noop {} {\bibfield  {journal} {\bibinfo  {journal}
  {\href{https://iopscience.iop.org/article/10.1088/1751-8113/47/42/424006/meta}{J.
  Phys. A: Math. Theor.}}\ }\textbf {\bibinfo {volume} {47}},\ \bibinfo {pages}
  {424006} (\bibinfo {year} {2014})}\BibitemShut {NoStop}%
\bibitem [{\citenamefont {Szczykulska}\ \emph {et~al.}(2016)\citenamefont
  {Szczykulska}, \citenamefont {Baumgratz},\ and\ \citenamefont
  {Datta}}]{szczykulska2016multi}%
  \BibitemOpen
  \bibfield  {author} {\bibinfo {author} {\bibfnamefont {M.}~\bibnamefont
  {Szczykulska}}, \bibinfo {author} {\bibfnamefont {T.}~\bibnamefont
  {Baumgratz}}, \ and\ \bibinfo {author} {\bibfnamefont {A.}~\bibnamefont
  {Datta}},\ }\href@noop {} {\bibfield  {journal} {\bibinfo  {journal}
  {\href{https://www.tandfonline.com/doi/full/10.1080/23746149.2016.1230476}{Adv.
  Phys. X}}\ }\textbf {\bibinfo {volume} {1}},\ \bibinfo {pages} {621}
  (\bibinfo {year} {2016})}\BibitemShut {NoStop}%
\bibitem [{\citenamefont {DeWitt}\ and\ \citenamefont
  {Esposito}(2008)}]{dewitt2008introduction}%
  \BibitemOpen
  \bibfield  {author} {\bibinfo {author} {\bibfnamefont {B.~S.}\ \bibnamefont
  {DeWitt}}\ and\ \bibinfo {author} {\bibfnamefont {G.}~\bibnamefont
  {Esposito}},\ }\href@noop {} {\bibfield  {journal} {\bibinfo  {journal}
  {\href{https://www.worldscientific.com/doi/abs/10.1142/S0219887808002679}{Int.
  J. Geom. Meth. Mod. Phys. }}\ }\textbf {\bibinfo {volume} {5}},\ \bibinfo
  {pages} {101} (\bibinfo {year} {2008})}\BibitemShut {NoStop}%
\bibitem [{\citenamefont {Takayanagi}(2012)}]{takayanagi2012entanglement}%
  \BibitemOpen
  \bibfield  {author} {\bibinfo {author} {\bibfnamefont {T.}~\bibnamefont
  {Takayanagi}},\ }\href@noop {} {\bibfield  {journal} {\bibinfo  {journal}
  {\href{https://iopscience.iop.org/article/10.1088/0264-9381/29/15/153001}{Class.
  Quantum Gravity}}\ }\textbf {\bibinfo {volume} {29}},\ \bibinfo {pages}
  {153001} (\bibinfo {year} {2012})}\BibitemShut {NoStop}%
\bibitem [{\citenamefont {Dong}(2014)}]{dong2014holographic}%
  \BibitemOpen
  \bibfield  {author} {\bibinfo {author} {\bibfnamefont {X.}~\bibnamefont
  {Dong}},\ }\href@noop {} {\bibfield  {journal} {\bibinfo  {journal}
  {\href{https://www.researchgate.net/publication/258083092_Holographic_Entanglement_Entropy_for_General_Higher_Derivative_Gravity}{J.
  High Energy Phys.}}\ }\textbf {\bibinfo {volume} {2014}},\ \bibinfo {pages}
  {44} (\bibinfo {year} {2014})}\BibitemShut {NoStop}%
\bibitem [{\citenamefont {Schulz}(2014)}]{schulz2014review}%
  \BibitemOpen
  \bibfield  {author} {\bibinfo {author} {\bibfnamefont {B.}~\bibnamefont
  {Schulz}},\ }\href@noop {} {\bibfield  {journal} {\bibinfo  {journal}
  {\href{https://arxiv.org/abs/1409.7977}{arXiv:1409.7977} [gr-qc]}\ }
  (\bibinfo {year} {2014})}\BibitemShut {NoStop}%
\bibitem [{\citenamefont {Nishioka}(2018)}]{nishioka2018entanglement}%
  \BibitemOpen
  \bibfield  {author} {\bibinfo {author} {\bibfnamefont {T.}~\bibnamefont
  {Nishioka}},\ }\href@noop {} {\bibfield  {journal} {\bibinfo  {journal}
  {\href{https://journals.aps.org/rmp/abstract/10.1103/RevModPhys.90.035007}{Rev.
  Mod. Phys.}}\ }\textbf {\bibinfo {volume} {90}},\ \bibinfo {pages} {035007}
  (\bibinfo {year} {2018})}\BibitemShut {NoStop}%
\bibitem [{\citenamefont {Helwig}\ \emph {et~al.}(2012)\citenamefont {Helwig},
  \citenamefont {Cui}, \citenamefont {Latorre}, \citenamefont {Riera},\ and\
  \citenamefont {Lo}}]{helwig2012absolute}%
  \BibitemOpen
  \bibfield  {author} {\bibinfo {author} {\bibfnamefont {W.}~\bibnamefont
  {Helwig}}, \bibinfo {author} {\bibfnamefont {W.}~\bibnamefont {Cui}},
  \bibinfo {author} {\bibfnamefont {J.~I.}\ \bibnamefont {Latorre}}, \bibinfo
  {author} {\bibfnamefont {A.}~\bibnamefont {Riera}}, \ and\ \bibinfo {author}
  {\bibfnamefont {H.-K.}\ \bibnamefont {Lo}},\ }\href@noop {} {\bibfield
  {journal} {\bibinfo  {journal}
  {\href{https://journals.aps.org/pra/abstract/10.1103/PhysRevA.86.052335}{Phys.
  Rev. A}}\ }\textbf {\bibinfo {volume} {86}},\ \bibinfo {pages} {052335}
  (\bibinfo {year} {2012})}\BibitemShut {NoStop}%
\bibitem [{\citenamefont {Bennett}\ and\ \citenamefont
  {Wiesner}(1992)}]{bennett1992communication}%
  \BibitemOpen
  \bibfield  {author} {\bibinfo {author} {\bibfnamefont {C.~H.}\ \bibnamefont
  {Bennett}}\ and\ \bibinfo {author} {\bibfnamefont {S.~J.}\ \bibnamefont
  {Wiesner}},\ }\href@noop {} {\bibfield  {journal} {\bibinfo  {journal}
  {\href{https://journals.aps.org/prl/abstract/10.1103/PhysRevLett.69.2881}{Phys.
  Rev. Lett.}}\ }\textbf {\bibinfo {volume} {69}},\ \bibinfo {pages} {2881}
  (\bibinfo {year} {1992})}\BibitemShut {NoStop}%
\bibitem [{\citenamefont {Zozulya}\ \emph {et~al.}(2007)\citenamefont
  {Zozulya}, \citenamefont {Haque}, \citenamefont {Schoutens},\ and\
  \citenamefont {Rezayi}}]{zozulya2007bipartite}%
  \BibitemOpen
  \bibfield  {author} {\bibinfo {author} {\bibfnamefont {O.~S.}\ \bibnamefont
  {Zozulya}}, \bibinfo {author} {\bibfnamefont {M.}~\bibnamefont {Haque}},
  \bibinfo {author} {\bibfnamefont {K.}~\bibnamefont {Schoutens}}, \ and\
  \bibinfo {author} {\bibfnamefont {E.~H.}\ \bibnamefont {Rezayi}},\
  }\href@noop {} {\bibfield  {journal} {\bibinfo  {journal}
  {\href{https://journals.aps.org/prb/abstract/10.1103/PhysRevB.76.125310}{Phys.
  Rev. B}}\ }\textbf {\bibinfo {volume} {76}},\ \bibinfo {pages} {125310}
  (\bibinfo {year} {2007})}\BibitemShut {NoStop}%
\bibitem [{\citenamefont {Li}\ and\ \citenamefont
  {Fei}(2010)}]{li2010measurable}%
  \BibitemOpen
  \bibfield  {author} {\bibinfo {author} {\bibfnamefont {M.}~\bibnamefont
  {Li}}\ and\ \bibinfo {author} {\bibfnamefont {S.-M.}\ \bibnamefont {Fei}},\
  }\href@noop {} {\bibfield  {journal} {\bibinfo  {journal}
  {\href{https://journals.aps.org/pra/abstract/10.1103/PhysRevA.82.044303}{Phys.
  Rev. A}}\ }\textbf {\bibinfo {volume} {82}},\ \bibinfo {pages} {044303}
  (\bibinfo {year} {2010})}\BibitemShut {NoStop}%
\bibitem [{\citenamefont {Jevtic}\ \emph {et~al.}(2012)\citenamefont {Jevtic},
  \citenamefont {Jennings},\ and\ \citenamefont
  {Rudolph}}]{jevtic2012maximally}%
  \BibitemOpen
  \bibfield  {author} {\bibinfo {author} {\bibfnamefont {S.}~\bibnamefont
  {Jevtic}}, \bibinfo {author} {\bibfnamefont {D.}~\bibnamefont {Jennings}}, \
  and\ \bibinfo {author} {\bibfnamefont {T.}~\bibnamefont {Rudolph}},\
  }\href@noop {} {\bibfield  {journal} {\bibinfo  {journal}
  {\href{https://journals.aps.org/prl/abstract/10.1103/PhysRevLett.108.110403}{Phys.
  Rev. Lett.}}\ }\textbf {\bibinfo {volume} {108}},\ \bibinfo {pages} {110403}
  (\bibinfo {year} {2012})}\BibitemShut {NoStop}%
\bibitem [{\citenamefont {Avery}\ and\ \citenamefont
  {Paulos}(2014)}]{avery2014universal}%
  \BibitemOpen
  \bibfield  {author} {\bibinfo {author} {\bibfnamefont {S.~G.}\ \bibnamefont
  {Avery}}\ and\ \bibinfo {author} {\bibfnamefont {M.~F.}\ \bibnamefont
  {Paulos}},\ }\href@noop {} {\bibfield  {journal} {\bibinfo  {journal}
  {\href{https://journals.aps.org/prl/abstract/10.1103/PhysRevLett.113.231604}{Phys.
  Rev. Lett.}}\ }\textbf {\bibinfo {volume} {113}},\ \bibinfo {pages} {231604}
  (\bibinfo {year} {2014})}\BibitemShut {NoStop}%
\bibitem [{\citenamefont {Song}\ \emph {et~al.}(2016)\citenamefont {Song},
  \citenamefont {Chen},\ and\ \citenamefont {Cao}}]{song2016lower}%
  \BibitemOpen
  \bibfield  {author} {\bibinfo {author} {\bibfnamefont {W.}~\bibnamefont
  {Song}}, \bibinfo {author} {\bibfnamefont {L.}~\bibnamefont {Chen}}, \ and\
  \bibinfo {author} {\bibfnamefont {Z.-L.}\ \bibnamefont {Cao}},\ }\href@noop
  {} {\bibfield  {journal} {\bibinfo  {journal}
  {\href{https://www.nature.com/articles/s41598-016-0029-9}{Sci. Rep.}}\
  }\textbf {\bibinfo {volume} {6}},\ \bibinfo {pages} {23} (\bibinfo {year}
  {2016})}\BibitemShut {NoStop}%
\bibitem [{\citenamefont {Vidmar}\ and\ \citenamefont
  {Rigol}(2017)}]{vidmar2017entanglement}%
  \BibitemOpen
  \bibfield  {author} {\bibinfo {author} {\bibfnamefont {L.}~\bibnamefont
  {Vidmar}}\ and\ \bibinfo {author} {\bibfnamefont {M.}~\bibnamefont {Rigol}},\
  }\href@noop {} {\bibfield  {journal} {\bibinfo  {journal}
  {\href{https://journals.aps.org/prl/abstract/10.1103/PhysRevLett.119.220603}{Phys.
  Rev. Lett.}}\ }\textbf {\bibinfo {volume} {119}},\ \bibinfo {pages} {220603}
  (\bibinfo {year} {2017})}\BibitemShut {NoStop}%
\bibitem [{\citenamefont {Vidmar}\ \emph {et~al.}(2017)\citenamefont {Vidmar},
  \citenamefont {Hackl}, \citenamefont {Bianchi},\ and\ \citenamefont
  {Rigol}}]{vidmar2017entanglementquadratic}%
  \BibitemOpen
  \bibfield  {author} {\bibinfo {author} {\bibfnamefont {L.}~\bibnamefont
  {Vidmar}}, \bibinfo {author} {\bibfnamefont {L.}~\bibnamefont {Hackl}},
  \bibinfo {author} {\bibfnamefont {E.}~\bibnamefont {Bianchi}}, \ and\
  \bibinfo {author} {\bibfnamefont {M.}~\bibnamefont {Rigol}},\ }\href@noop {}
  {\bibfield  {journal} {\bibinfo  {journal}
  {\href{https://journals.aps.org/prl/abstract/10.1103/PhysRevLett.119.020601}{Phys.
  Rev. Lett.}}\ }\textbf {\bibinfo {volume} {119}},\ \bibinfo {pages} {020601}
  (\bibinfo {year} {2017})}\BibitemShut {NoStop}%
\bibitem [{\citenamefont {Beaud}\ and\ \citenamefont
  {Warzel}(2018)}]{beaud2018bounds}%
  \BibitemOpen
  \bibfield  {author} {\bibinfo {author} {\bibfnamefont {V.}~\bibnamefont
  {Beaud}}\ and\ \bibinfo {author} {\bibfnamefont {S.}~\bibnamefont {Warzel}},\
  }\href@noop {} {\bibfield  {journal} {\bibinfo  {journal}
  {\href{https://aip.scitation.org/doi/10.1063/1.5007035}{J. Math. Phys.}}\
  }\textbf {\bibinfo {volume} {59}},\ \bibinfo {pages} {012109} (\bibinfo
  {year} {2018})}\BibitemShut {NoStop}%
\bibitem [{\citenamefont {B{\"a}uml}\ \emph {et~al.}(2018)\citenamefont
  {B{\"a}uml}, \citenamefont {Das},\ and\ \citenamefont
  {Wilde}}]{bauml2018fundamental}%
  \BibitemOpen
  \bibfield  {author} {\bibinfo {author} {\bibfnamefont {S.}~\bibnamefont
  {B{\"a}uml}}, \bibinfo {author} {\bibfnamefont {S.}~\bibnamefont {Das}}, \
  and\ \bibinfo {author} {\bibfnamefont {M.~M.}\ \bibnamefont {Wilde}},\
  }\href@noop {} {\bibfield  {journal} {\bibinfo  {journal}
  {\href{https://journals.aps.org/prl/abstract/10.1103/PhysRevLett.121.250504}{Phys.
  Rev. Lett.}}\ }\textbf {\bibinfo {volume} {121}},\ \bibinfo {pages} {250504}
  (\bibinfo {year} {2018})}\BibitemShut {NoStop}%
\bibitem [{\citenamefont {Fujita}\ \emph {et~al.}(2018)\citenamefont {Fujita},
  \citenamefont {Nakagawa}, \citenamefont {Sugiura},\ and\ \citenamefont
  {Watanabe}}]{fujita2018page}%
  \BibitemOpen
  \bibfield  {author} {\bibinfo {author} {\bibfnamefont {H.}~\bibnamefont
  {Fujita}}, \bibinfo {author} {\bibfnamefont {Y.~O.}\ \bibnamefont
  {Nakagawa}}, \bibinfo {author} {\bibfnamefont {S.}~\bibnamefont {Sugiura}}, \
  and\ \bibinfo {author} {\bibfnamefont {M.}~\bibnamefont {Watanabe}},\
  }\href@noop {} {\bibfield  {journal} {\bibinfo  {journal}
  {\href{https://link.springer.com/article/10.1007/JHEP12(2018)112}{J. High
  Energy Phys.}}\ }\textbf {\bibinfo {volume} {2018}},\ \bibinfo {pages} {112}
  (\bibinfo {year} {2018})}\BibitemShut {NoStop}%
\bibitem [{\citenamefont {Huang}(2019{\natexlab{a}})}]{huang2019universal}%
  \BibitemOpen
  \bibfield  {author} {\bibinfo {author} {\bibfnamefont {Y.}~\bibnamefont
  {Huang}},\ }\href@noop {} {\bibfield  {journal} {\bibinfo  {journal}
  {\href{https://doi.org/10.1016/j.nuclphysb.2018.09.013}{Nucl. Phys. B}}\
  }\textbf {\bibinfo {volume} {938}},\ \bibinfo {pages} {594} (\bibinfo {year}
  {2019}{\natexlab{a}})}\BibitemShut {NoStop}%
\bibitem [{\citenamefont {Huang}(2019{\natexlab{b}})}]{huang2019dynamics}%
  \BibitemOpen
  \bibfield  {author} {\bibinfo {author} {\bibfnamefont {Y.}~\bibnamefont
  {Huang}},\ }\href@noop {} {\bibfield  {journal} {\bibinfo  {journal}
  {\href{https://arxiv.org/abs/1902.00977}{arXiv:1902.00977} [quant-ph]}\ }
  (\bibinfo {year} {2019}{\natexlab{b}})}\BibitemShut {NoStop}%
\bibitem [{\citenamefont {Wiseman}\ and\ \citenamefont
  {Vaccaro}(2003)}]{wiseman2003entanglement}%
  \BibitemOpen
  \bibfield  {author} {\bibinfo {author} {\bibfnamefont {H.~M.}\ \bibnamefont
  {Wiseman}}\ and\ \bibinfo {author} {\bibfnamefont {J.~A.}\ \bibnamefont
  {Vaccaro}},\ }\href@noop {} {\bibfield  {journal} {\bibinfo  {journal}
  {\href{https://journals.aps.org/prl/abstract/10.1103/PhysRevLett.91.097902}{Phys.
  Rev. Lett.}}\ }\textbf {\bibinfo {volume} {91}},\ \bibinfo {pages} {097902}
  (\bibinfo {year} {2003})}\BibitemShut {NoStop}%
\bibitem [{\citenamefont {Bennett}\ \emph {et~al.}(2011)\citenamefont
  {Bennett}, \citenamefont {Grudka}, \citenamefont {Horodecki}, \citenamefont
  {Horodecki},\ and\ \citenamefont {Horodecki}}]{bennett2011postulates}%
  \BibitemOpen
  \bibfield  {author} {\bibinfo {author} {\bibfnamefont {C.~H.}\ \bibnamefont
  {Bennett}}, \bibinfo {author} {\bibfnamefont {A.}~\bibnamefont {Grudka}},
  \bibinfo {author} {\bibfnamefont {M.}~\bibnamefont {Horodecki}}, \bibinfo
  {author} {\bibfnamefont {P.}~\bibnamefont {Horodecki}}, \ and\ \bibinfo
  {author} {\bibfnamefont {R.}~\bibnamefont {Horodecki}},\ }\href@noop {}
  {\bibfield  {journal} {\bibinfo  {journal}
  {\href{https://journals.aps.org/pra/abstract/10.1103/PhysRevA.83.012312}{Phys.
  Rev. A}}\ }\textbf {\bibinfo {volume} {83}},\ \bibinfo {pages} {012312}
  (\bibinfo {year} {2011})}\BibitemShut {NoStop}%
\bibitem [{\citenamefont {Plenio}\ and\ \citenamefont
  {Virmani}(2011)}]{plenio2014introduction}%
  \BibitemOpen
  \bibfield  {author} {\bibinfo {author} {\bibfnamefont {M.~B.}\ \bibnamefont
  {Plenio}}\ and\ \bibinfo {author} {\bibfnamefont {S.~S.}\ \bibnamefont
  {Virmani}},\ }\enquote {\bibinfo {title}
  {\href{https://arxiv.org/pdf/quant-ph/0504163.pdf}{An introduction to
  entanglement theory}},}\ in\ \href@noop {} {\emph {\bibinfo {booktitle}
  {Quantum information and coherence}}}\ (\bibinfo  {publisher} {Springer},\
  \bibinfo {year} {2011})\ pp.\ \bibinfo {pages} {173--209}\BibitemShut
  {NoStop}%
\bibitem [{\citenamefont {Girolami}\ \emph {et~al.}(2017)\citenamefont
  {Girolami}, \citenamefont {Tufarelli},\ and\ \citenamefont
  {Susa}}]{girolami2017quantifying}%
  \BibitemOpen
  \bibfield  {author} {\bibinfo {author} {\bibfnamefont {D.}~\bibnamefont
  {Girolami}}, \bibinfo {author} {\bibfnamefont {T.}~\bibnamefont {Tufarelli}},
  \ and\ \bibinfo {author} {\bibfnamefont {C.~E.}\ \bibnamefont {Susa}},\
  }\href@noop {} {\bibfield  {journal} {\bibinfo  {journal}
  {\href{https://journals.aps.org/prl/abstract/10.1103/PhysRevLett.119.140505}{Phys.
  Rev. Lett.}}\ }\textbf {\bibinfo {volume} {119}},\ \bibinfo {pages} {140505}
  (\bibinfo {year} {2017})}\BibitemShut {NoStop}%
\bibitem [{\citenamefont {Page}(1993)}]{page1993average}%
  \BibitemOpen
  \bibfield  {author} {\bibinfo {author} {\bibfnamefont {D.~N.}\ \bibnamefont
  {Page}},\ }\href@noop {} {\bibfield  {journal} {\bibinfo  {journal}
  {\href{https://journals.aps.org/prl/abstract/10.1103/PhysRevLett.71.1291}{Phys.
  Rev. Lett.}}\ }\textbf {\bibinfo {volume} {71}},\ \bibinfo {pages} {1291}
  (\bibinfo {year} {1993})}\BibitemShut {NoStop}%
\bibitem [{\citenamefont {Nakagawa}\ \emph {et~al.}(2018)\citenamefont
  {Nakagawa}, \citenamefont {Watanabe}, \citenamefont {Fujita},\ and\
  \citenamefont {Sugiura}}]{nakagawa2018universality}%
  \BibitemOpen
  \bibfield  {author} {\bibinfo {author} {\bibfnamefont {Y.~O.}\ \bibnamefont
  {Nakagawa}}, \bibinfo {author} {\bibfnamefont {M.}~\bibnamefont {Watanabe}},
  \bibinfo {author} {\bibfnamefont {H.}~\bibnamefont {Fujita}}, \ and\ \bibinfo
  {author} {\bibfnamefont {S.}~\bibnamefont {Sugiura}},\ }\href@noop {}
  {\bibfield  {journal} {\bibinfo  {journal}
  {\href{https://www.nature.com/articles/s41467-018-03883-9}{Nat. Comm.}}\
  }\textbf {\bibinfo {volume} {9}},\ \bibinfo {pages} {1635} (\bibinfo {year}
  {2018})}\BibitemShut {NoStop}%
\bibitem [{\citenamefont {Eisert}\ \emph {et~al.}(2010)\citenamefont {Eisert},
  \citenamefont {Cramer},\ and\ \citenamefont {Plenio}}]{eisert2008area}%
  \BibitemOpen
  \bibfield  {author} {\bibinfo {author} {\bibfnamefont {J.}~\bibnamefont
  {Eisert}}, \bibinfo {author} {\bibfnamefont {M.}~\bibnamefont {Cramer}}, \
  and\ \bibinfo {author} {\bibfnamefont {M.~B.}\ \bibnamefont {Plenio}},\
  }\href@noop {} {\bibfield  {journal} {\bibinfo  {journal}
  {\href{https://journals.aps.org/rmp/abstract/10.1103/RevModPhys.82.277}{Rev.
  Mod. Phys.}}\ }\textbf {\bibinfo {volume} {82}},\ \bibinfo {pages} {277}
  (\bibinfo {year} {2010})}\BibitemShut {NoStop}%
\bibitem [{\citenamefont {Laflorencie}(2016)}]{laflorencie2016quantum}%
  \BibitemOpen
  \bibfield  {author} {\bibinfo {author} {\bibfnamefont {N.}~\bibnamefont
  {Laflorencie}},\ }\href@noop {} {\bibfield  {journal} {\bibinfo  {journal}
  {\href{https://www.sciencedirect.com/science/article/abs/pii/S0370157316301582}{Phys.
  Rep.}}\ }\textbf {\bibinfo {volume} {646}},\ \bibinfo {pages} {1} (\bibinfo
  {year} {2016})}\BibitemShut {NoStop}%
\bibitem [{\citenamefont {Cho}(2018)}]{cho2018realistic}%
  \BibitemOpen
  \bibfield  {author} {\bibinfo {author} {\bibfnamefont {J.}~\bibnamefont
  {Cho}},\ }\href@noop {} {\bibfield  {journal} {\bibinfo  {journal}
  {\href{https://journals.aps.org/prx/abstract/10.1103/PhysRevX.8.031009}{Phys.
  Rev. X}}\ }\textbf {\bibinfo {volume} {8}},\ \bibinfo {pages} {031009}
  (\bibinfo {year} {2018})}\BibitemShut {NoStop}%
\bibitem [{\citenamefont {Gu}\ \emph {et~al.}(2004)\citenamefont {Gu},
  \citenamefont {Deng}, \citenamefont {Li},\ and\ \citenamefont
  {Lin}}]{gu2004entanglement}%
  \BibitemOpen
  \bibfield  {author} {\bibinfo {author} {\bibfnamefont {S.-J.}\ \bibnamefont
  {Gu}}, \bibinfo {author} {\bibfnamefont {S.-S.}\ \bibnamefont {Deng}},
  \bibinfo {author} {\bibfnamefont {Y.-Q.}\ \bibnamefont {Li}}, \ and\ \bibinfo
  {author} {\bibfnamefont {H.-Q.}\ \bibnamefont {Lin}},\ }\href@noop {}
  {\bibfield  {journal} {\bibinfo  {journal}
  {\href{https://journals.aps.org/prl/abstract/10.1103/PhysRevLett.93.086402}{Phys.
  Rev. Lett.}}\ }\textbf {\bibinfo {volume} {93}},\ \bibinfo {pages} {086402}
  (\bibinfo {year} {2004})}\BibitemShut {NoStop}%
\bibitem [{\citenamefont {Le~Hur}(2008)}]{le2008entanglement}%
  \BibitemOpen
  \bibfield  {author} {\bibinfo {author} {\bibfnamefont {K.}~\bibnamefont
  {Le~Hur}},\ }\href@noop {} {\bibfield  {journal} {\bibinfo  {journal} {Annals
  of Physics}\ }\textbf {\bibinfo {volume} {323}},\ \bibinfo {pages} {2208}
  (\bibinfo {year} {2008})}\BibitemShut {NoStop}%
\bibitem [{\citenamefont {Barghathi}\ \emph {et~al.}(2019)\citenamefont
  {Barghathi}, \citenamefont {Casiano-Diaz},\ and\ \citenamefont
  {Del~Maestro}}]{barghathi2019operationally}%
  \BibitemOpen
  \bibfield  {author} {\bibinfo {author} {\bibfnamefont {H.}~\bibnamefont
  {Barghathi}}, \bibinfo {author} {\bibfnamefont {E.}~\bibnamefont
  {Casiano-Diaz}}, \ and\ \bibinfo {author} {\bibfnamefont {A.}~\bibnamefont
  {Del~Maestro}},\ }\href@noop {} {\bibfield  {journal} {\bibinfo  {journal}
  {\href{https://arxiv.org/abs/1905.03312}{arXiv:1905.03312} [quant-ph]}\ }
  (\bibinfo {year} {2019})}\BibitemShut {NoStop}%
\bibitem [{\citenamefont {Islam}\ \emph {et~al.}(2015)\citenamefont {Islam},
  \citenamefont {Ma}, \citenamefont {Preiss}, \citenamefont {Tai},
  \citenamefont {Lukin}, \citenamefont {Rispoli},\ and\ \citenamefont
  {Greiner}}]{islam2015measuring}%
  \BibitemOpen
  \bibfield  {author} {\bibinfo {author} {\bibfnamefont {R.}~\bibnamefont
  {Islam}}, \bibinfo {author} {\bibfnamefont {R.}~\bibnamefont {Ma}}, \bibinfo
  {author} {\bibfnamefont {P.~M.}\ \bibnamefont {Preiss}}, \bibinfo {author}
  {\bibfnamefont {M.~E.}\ \bibnamefont {Tai}}, \bibinfo {author} {\bibfnamefont
  {A.}~\bibnamefont {Lukin}}, \bibinfo {author} {\bibfnamefont
  {M.}~\bibnamefont {Rispoli}}, \ and\ \bibinfo {author} {\bibfnamefont
  {M.}~\bibnamefont {Greiner}},\ }\href@noop {} {\bibfield  {journal} {\bibinfo
   {journal} {\href{https://www.nature.com/articles/nature15750}{Nature}}\
  }\textbf {\bibinfo {volume} {528}},\ \bibinfo {pages} {77} (\bibinfo {year}
  {2015})}\BibitemShut {NoStop}%
\bibitem [{\citenamefont {Kaufman}\ \emph {et~al.}(2016)\citenamefont
  {Kaufman}, \citenamefont {Tai}, \citenamefont {Lukin}, \citenamefont
  {Rispoli}, \citenamefont {Schittko}, \citenamefont {Preiss},\ and\
  \citenamefont {Greiner}}]{kaufman2016quantum}%
  \BibitemOpen
  \bibfield  {author} {\bibinfo {author} {\bibfnamefont {A.~M.}\ \bibnamefont
  {Kaufman}}, \bibinfo {author} {\bibfnamefont {M.~E.}\ \bibnamefont {Tai}},
  \bibinfo {author} {\bibfnamefont {A.}~\bibnamefont {Lukin}}, \bibinfo
  {author} {\bibfnamefont {M.}~\bibnamefont {Rispoli}}, \bibinfo {author}
  {\bibfnamefont {R.}~\bibnamefont {Schittko}}, \bibinfo {author}
  {\bibfnamefont {P.~M.}\ \bibnamefont {Preiss}}, \ and\ \bibinfo {author}
  {\bibfnamefont {M.}~\bibnamefont {Greiner}},\ }\href@noop {} {\bibfield
  {journal} {\bibinfo  {journal}
  {\href{https://science.sciencemag.org/content/353/6301/794}{Science}}\
  }\textbf {\bibinfo {volume} {353}},\ \bibinfo {pages} {794} (\bibinfo {year}
  {2016})}\BibitemShut {NoStop}%
\bibitem [{\citenamefont {Brydges}\ \emph {et~al.}(2019)\citenamefont
  {Brydges}, \citenamefont {Elben}, \citenamefont {Jurcevic}, \citenamefont
  {Vermersch}, \citenamefont {Maier}, \citenamefont {Lanyon}, \citenamefont
  {Zoller}, \citenamefont {Blatt},\ and\ \citenamefont
  {Roos}}]{brydges2018probing}%
  \BibitemOpen
  \bibfield  {author} {\bibinfo {author} {\bibfnamefont {T.}~\bibnamefont
  {Brydges}}, \bibinfo {author} {\bibfnamefont {A.}~\bibnamefont {Elben}},
  \bibinfo {author} {\bibfnamefont {P.}~\bibnamefont {Jurcevic}}, \bibinfo
  {author} {\bibfnamefont {B.}~\bibnamefont {Vermersch}}, \bibinfo {author}
  {\bibfnamefont {C.}~\bibnamefont {Maier}}, \bibinfo {author} {\bibfnamefont
  {B.~P.}\ \bibnamefont {Lanyon}}, \bibinfo {author} {\bibfnamefont
  {P.}~\bibnamefont {Zoller}}, \bibinfo {author} {\bibfnamefont
  {R.}~\bibnamefont {Blatt}}, \ and\ \bibinfo {author} {\bibfnamefont {C.~F.}\
  \bibnamefont {Roos}},\ }\href@noop {} {\bibfield  {journal} {\bibinfo
  {journal}
  {\href{https://science.sciencemag.org/content/364/6437/260}{Science}}\
  }\textbf {\bibinfo {volume} {364}},\ \bibinfo {pages} {260} (\bibinfo {year}
  {2019})}\BibitemShut {NoStop}%
\bibitem [{\citenamefont {Mendes-Santos}\ \emph {et~al.}(2019)\citenamefont
  {Mendes-Santos}, \citenamefont {Giudici}, \citenamefont {Fazio},\ and\
  \citenamefont {Dalmonte}}]{mendes2019measuring}%
  \BibitemOpen
  \bibfield  {author} {\bibinfo {author} {\bibfnamefont {T.}~\bibnamefont
  {Mendes-Santos}}, \bibinfo {author} {\bibfnamefont {G.}~\bibnamefont
  {Giudici}}, \bibinfo {author} {\bibfnamefont {R.}~\bibnamefont {Fazio}}, \
  and\ \bibinfo {author} {\bibfnamefont {M.}~\bibnamefont {Dalmonte}},\
  }\href@noop {} {\bibfield  {journal} {\bibinfo  {journal}
  {\href{https://arxiv.org/abs/1904.07782}{arXiv:1904.07782}
  [cond-mat.str-el]}\ } (\bibinfo {year} {2019})}\BibitemShut {NoStop}%
\bibitem [{\citenamefont {Daley}\ \emph {et~al.}(2012)\citenamefont {Daley},
  \citenamefont {Pichler}, \citenamefont {Schachenmayer},\ and\ \citenamefont
  {Zoller}}]{daley2012measuring}%
  \BibitemOpen
  \bibfield  {author} {\bibinfo {author} {\bibfnamefont {A.~J.}\ \bibnamefont
  {Daley}}, \bibinfo {author} {\bibfnamefont {H.}~\bibnamefont {Pichler}},
  \bibinfo {author} {\bibfnamefont {J.}~\bibnamefont {Schachenmayer}}, \ and\
  \bibinfo {author} {\bibfnamefont {P.}~\bibnamefont {Zoller}},\ }\href@noop {}
  {\bibfield  {journal} {\bibinfo  {journal}
  {\href{https://journals.aps.org/prl/abstract/10.1103/PhysRevLett.109.020505}{Phys.
  Rev. Lett.}}\ }\textbf {\bibinfo {volume} {109}},\ \bibinfo {pages} {020505}
  (\bibinfo {year} {2012})}\BibitemShut {NoStop}%
\bibitem [{\citenamefont {Pichler}\ \emph {et~al.}(2016)\citenamefont
  {Pichler}, \citenamefont {Zhu}, \citenamefont {Seif}, \citenamefont
  {Zoller},\ and\ \citenamefont {Hafezi}}]{pichler2016measurement}%
  \BibitemOpen
  \bibfield  {author} {\bibinfo {author} {\bibfnamefont {H.}~\bibnamefont
  {Pichler}}, \bibinfo {author} {\bibfnamefont {G.}~\bibnamefont {Zhu}},
  \bibinfo {author} {\bibfnamefont {A.}~\bibnamefont {Seif}}, \bibinfo {author}
  {\bibfnamefont {P.}~\bibnamefont {Zoller}}, \ and\ \bibinfo {author}
  {\bibfnamefont {M.}~\bibnamefont {Hafezi}},\ }\href@noop {} {\bibfield
  {journal} {\bibinfo  {journal}
  {\href{https://journals.aps.org/prx/abstract/10.1103/PhysRevX.6.041033}{Phys.
  Rev. X}}\ }\textbf {\bibinfo {volume} {6}},\ \bibinfo {pages} {041033}
  (\bibinfo {year} {2016})}\BibitemShut {NoStop}%
\bibitem [{\citenamefont {Barghathi}\ \emph {et~al.}(2018)\citenamefont
  {Barghathi}, \citenamefont {Herdman},\ and\ \citenamefont
  {Del~Maestro}}]{barghathi2018r}%
  \BibitemOpen
  \bibfield  {author} {\bibinfo {author} {\bibfnamefont {H.}~\bibnamefont
  {Barghathi}}, \bibinfo {author} {\bibfnamefont {C.~M.}\ \bibnamefont
  {Herdman}}, \ and\ \bibinfo {author} {\bibfnamefont {A.}~\bibnamefont
  {Del~Maestro}},\ }\href@noop {} {\bibfield  {journal} {\bibinfo  {journal}
  {\href{https://journals.aps.org/prl/abstract/10.1103/PhysRevLett.121.150501}{Phys.
  Rev. Lett.}}\ }\textbf {\bibinfo {volume} {121}},\ \bibinfo {pages} {150501}
  (\bibinfo {year} {2018})}\BibitemShut {NoStop}%
\bibitem [{\citenamefont {Linke}\ \emph {et~al.}(2018)\citenamefont {Linke},
  \citenamefont {Johri}, \citenamefont {Figgatt}, \citenamefont {Landsman},
  \citenamefont {Matsuura},\ and\ \citenamefont {Monroe}}]{linke2018measuring}%
  \BibitemOpen
  \bibfield  {author} {\bibinfo {author} {\bibfnamefont {N.~M.}\ \bibnamefont
  {Linke}}, \bibinfo {author} {\bibfnamefont {S.}~\bibnamefont {Johri}},
  \bibinfo {author} {\bibfnamefont {C.}~\bibnamefont {Figgatt}}, \bibinfo
  {author} {\bibfnamefont {K.~A.}\ \bibnamefont {Landsman}}, \bibinfo {author}
  {\bibfnamefont {A.~Y.}\ \bibnamefont {Matsuura}}, \ and\ \bibinfo {author}
  {\bibfnamefont {C.}~\bibnamefont {Monroe}},\ }\href@noop {} {\bibfield
  {journal} {\bibinfo  {journal}
  {\href{https://journals.aps.org/pra/abstract/10.1103/PhysRevA.98.052334}{Phys.
  Rev. A}}\ }\textbf {\bibinfo {volume} {98}},\ \bibinfo {pages} {052334}
  (\bibinfo {year} {2018})}\BibitemShut {NoStop}%
\bibitem [{\citenamefont {Lukin}\ \emph {et~al.}(2019)\citenamefont {Lukin},
  \citenamefont {Rispoli}, \citenamefont {Schittko}, \citenamefont {Tai},
  \citenamefont {Kaufman}, \citenamefont {Choi}, \citenamefont {Khemani},
  \citenamefont {L{\'e}onard},\ and\ \citenamefont
  {Greiner}}]{lukin2019probing}%
  \BibitemOpen
  \bibfield  {author} {\bibinfo {author} {\bibfnamefont {A.}~\bibnamefont
  {Lukin}}, \bibinfo {author} {\bibfnamefont {M.}~\bibnamefont {Rispoli}},
  \bibinfo {author} {\bibfnamefont {R.}~\bibnamefont {Schittko}}, \bibinfo
  {author} {\bibfnamefont {M.~E.}\ \bibnamefont {Tai}}, \bibinfo {author}
  {\bibfnamefont {A.~M.}\ \bibnamefont {Kaufman}}, \bibinfo {author}
  {\bibfnamefont {S.}~\bibnamefont {Choi}}, \bibinfo {author} {\bibfnamefont
  {V.}~\bibnamefont {Khemani}}, \bibinfo {author} {\bibfnamefont
  {J.}~\bibnamefont {L{\'e}onard}}, \ and\ \bibinfo {author} {\bibfnamefont
  {M.}~\bibnamefont {Greiner}},\ }\href@noop {} {\bibfield  {journal} {\bibinfo
   {journal}
  {\href{https://science.sciencemag.org/content/364/6437/256}{Science}}\
  }\textbf {\bibinfo {volume} {364}},\ \bibinfo {pages} {256} (\bibinfo {year}
  {2019})}\BibitemShut {NoStop}%
\bibitem [{\citenamefont {Bartlett}\ \emph {et~al.}(2007)\citenamefont
  {Bartlett}, \citenamefont {Rudolph},\ and\ \citenamefont
  {Spekkens}}]{bartlett2007reference}%
  \BibitemOpen
  \bibfield  {author} {\bibinfo {author} {\bibfnamefont {S.~D.}\ \bibnamefont
  {Bartlett}}, \bibinfo {author} {\bibfnamefont {T.}~\bibnamefont {Rudolph}}, \
  and\ \bibinfo {author} {\bibfnamefont {R.~W.}\ \bibnamefont {Spekkens}},\
  }\href@noop {} {\bibfield  {journal} {\bibinfo  {journal}
  {\href{https://journals.aps.org/rmp/abstract/10.1103/RevModPhys.79.555}{Rev.
  Mod. Phys.}}\ }\textbf {\bibinfo {volume} {79}},\ \bibinfo {pages} {555}
  (\bibinfo {year} {2007})}\BibitemShut {NoStop}%
\bibitem [{\citenamefont {Benatti}\ \emph {et~al.}(2012)\citenamefont
  {Benatti}, \citenamefont {Floreanini},\ and\ \citenamefont
  {Marzolino}}]{benatti2012bipartite}%
  \BibitemOpen
  \bibfield  {author} {\bibinfo {author} {\bibfnamefont {F.}~\bibnamefont
  {Benatti}}, \bibinfo {author} {\bibfnamefont {R.}~\bibnamefont {Floreanini}},
  \ and\ \bibinfo {author} {\bibfnamefont {U.}~\bibnamefont {Marzolino}},\
  }\href@noop {} {\bibfield  {journal} {\bibinfo  {journal}
  {\href{https://www.sciencedirect.com/science/article/pii/S0003491612000206}{Ann.
  Phys.}}\ }\textbf {\bibinfo {volume} {327}},\ \bibinfo {pages} {1304}
  (\bibinfo {year} {2012})}\BibitemShut {NoStop}%
\bibitem [{\citenamefont {Ionicioiu}\ and\ \citenamefont
  {Zanardi}(2002)}]{ionicioiu2002quantum}%
  \BibitemOpen
  \bibfield  {author} {\bibinfo {author} {\bibfnamefont {R.}~\bibnamefont
  {Ionicioiu}}\ and\ \bibinfo {author} {\bibfnamefont {P.}~\bibnamefont
  {Zanardi}},\ }\href@noop {} {\bibfield  {journal} {\bibinfo  {journal}
  {\href{https://journals.aps.org/pra/abstract/10.1103/PhysRevA.66.050301}{Phys.
  Rev. A}}\ }\textbf {\bibinfo {volume} {66}},\ \bibinfo {pages} {050301(R)}
  (\bibinfo {year} {2002})}\BibitemShut {NoStop}%
\bibitem [{\citenamefont {Verstraete}\ and\ \citenamefont
  {Cirac}(2003)}]{verstraete2003quantum}%
  \BibitemOpen
  \bibfield  {author} {\bibinfo {author} {\bibfnamefont {F.}~\bibnamefont
  {Verstraete}}\ and\ \bibinfo {author} {\bibfnamefont {J.~I.}\ \bibnamefont
  {Cirac}},\ }\href@noop {} {\bibfield  {journal} {\bibinfo  {journal}
  {\href{https://journals.aps.org/prl/abstract/10.1103/PhysRevLett.91.010404}{Phys.
  Rev. Lett.}}\ }\textbf {\bibinfo {volume} {91}},\ \bibinfo {pages} {010404}
  (\bibinfo {year} {2003})}\BibitemShut {NoStop}%
\bibitem [{\citenamefont {Marzolino}\ and\ \citenamefont
  {Buchleitner}(2015)}]{marzolino2015quantum}%
  \BibitemOpen
  \bibfield  {author} {\bibinfo {author} {\bibfnamefont {U.}~\bibnamefont
  {Marzolino}}\ and\ \bibinfo {author} {\bibfnamefont {A.}~\bibnamefont
  {Buchleitner}},\ }\href@noop {} {\bibfield  {journal} {\bibinfo  {journal}
  {\href{https://journals.aps.org/pra/abstract/10.1103/PhysRevA.91.032316}{Phys.
  Rev. A}}\ }\textbf {\bibinfo {volume} {91}},\ \bibinfo {pages} {032316}
  (\bibinfo {year} {2015})}\BibitemShut {NoStop}%
\bibitem [{\citenamefont {Marzolino}\ and\ \citenamefont
  {Buchleitner}(2016)}]{marzolino2016performances}%
  \BibitemOpen
  \bibfield  {author} {\bibinfo {author} {\bibfnamefont {U.}~\bibnamefont
  {Marzolino}}\ and\ \bibinfo {author} {\bibfnamefont {A.}~\bibnamefont
  {Buchleitner}},\ }\href@noop {} {\bibfield  {journal} {\bibinfo  {journal}
  {\href{https://royalsocietypublishing.org/doi/10.1098/rspa.2015.0621}{ Proc.
  R. Soc. A }}\ }\textbf {\bibinfo {volume} {472}},\ \bibinfo {pages}
  {20150621} (\bibinfo {year} {2016})}\BibitemShut {NoStop}%
\bibitem [{\citenamefont {Benatti}\ \emph {et~al.}(2010)\citenamefont
  {Benatti}, \citenamefont {Floreanini},\ and\ \citenamefont
  {Marzolino}}]{benatti2010sub}%
  \BibitemOpen
  \bibfield  {author} {\bibinfo {author} {\bibfnamefont {F.}~\bibnamefont
  {Benatti}}, \bibinfo {author} {\bibfnamefont {R.}~\bibnamefont {Floreanini}},
  \ and\ \bibinfo {author} {\bibfnamefont {U.}~\bibnamefont {Marzolino}},\
  }\href@noop {} {\bibfield  {journal} {\bibinfo  {journal}
  {\href{https://www.sciencedirect.com/science/article/pii/S0003491610000084}{Ann.
  Phys.}}\ }\textbf {\bibinfo {volume} {325}},\ \bibinfo {pages} {924}
  (\bibinfo {year} {2010})}\BibitemShut {NoStop}%
\bibitem [{\citenamefont {Gross}\ \emph {et~al.}(2010)\citenamefont {Gross},
  \citenamefont {Zibold}, \citenamefont {Nicklas}, \citenamefont {Esteve},\
  and\ \citenamefont {Oberthaler}}]{gross2010nonlinear}%
  \BibitemOpen
  \bibfield  {author} {\bibinfo {author} {\bibfnamefont {C.}~\bibnamefont
  {Gross}}, \bibinfo {author} {\bibfnamefont {T.}~\bibnamefont {Zibold}},
  \bibinfo {author} {\bibfnamefont {E.}~\bibnamefont {Nicklas}}, \bibinfo
  {author} {\bibfnamefont {J.}~\bibnamefont {Esteve}}, \ and\ \bibinfo {author}
  {\bibfnamefont {M.~K.}\ \bibnamefont {Oberthaler}},\ }\href@noop {}
  {\bibfield  {journal} {\bibinfo  {journal}
  {\href{https://www.nature.com/articles/nature08919}{Nature}}\ }\textbf
  {\bibinfo {volume} {464}},\ \bibinfo {pages} {1165} (\bibinfo {year}
  {2010})}\BibitemShut {NoStop}%
\bibitem [{\citenamefont {Benatti}\ \emph {et~al.}(2011)\citenamefont
  {Benatti}, \citenamefont {Floreanini},\ and\ \citenamefont
  {Marzolino}}]{benatti2011entanglement}%
  \BibitemOpen
  \bibfield  {author} {\bibinfo {author} {\bibfnamefont {F.}~\bibnamefont
  {Benatti}}, \bibinfo {author} {\bibfnamefont {R.}~\bibnamefont {Floreanini}},
  \ and\ \bibinfo {author} {\bibfnamefont {U.}~\bibnamefont {Marzolino}},\
  }\href@noop {} {\bibfield  {journal} {\bibinfo  {journal}
  {\href{https://iopscience.iop.org/article/10.1088/0953-4075/44/9/091001/meta}{J.
  Phys. B}}\ }\textbf {\bibinfo {volume} {44}},\ \bibinfo {pages} {091001}
  (\bibinfo {year} {2011})}\BibitemShut {NoStop}%
\bibitem [{\citenamefont {Benatti}\ \emph {et~al.}(2014)\citenamefont
  {Benatti}, \citenamefont {Floreanini},\ and\ \citenamefont
  {Marzolino}}]{benatti2014entanglement}%
  \BibitemOpen
  \bibfield  {author} {\bibinfo {author} {\bibfnamefont {F.}~\bibnamefont
  {Benatti}}, \bibinfo {author} {\bibfnamefont {R.}~\bibnamefont {Floreanini}},
  \ and\ \bibinfo {author} {\bibfnamefont {U.}~\bibnamefont {Marzolino}},\
  }\href@noop {} {\bibfield  {journal} {\bibinfo  {journal}
  {\href{https://journals.aps.org/pra/abstract/10.1103/PhysRevA.89.032326}{Phys.
  Rev. A}}\ }\textbf {\bibinfo {volume} {89}},\ \bibinfo {pages} {032326}
  (\bibinfo {year} {2014})}\BibitemShut {NoStop}%
\bibitem [{Note1()}]{Note1}%
  \BibitemOpen
  \bibinfo {note} {This is necessary, but not a sufficient condition for the
  state to be maximally entangled. This means that even if the Eqs.~\protect
  \textup {\hbox {\mathsurround \z@ \protect \normalfont (\ignorespaces \ref
  {eq:fermprob}\unskip \@@italiccorr )}} or \protect \textup {\hbox
  {\mathsurround \z@ \protect \normalfont (\ignorespaces \ref
  {eq:bosprob}\unskip \@@italiccorr )}} are satisfied, the state does not have
  to be maximally entangled. The sufficient and necessary condition is given by
  Eq.~\protect \textup {\hbox {\mathsurround \z@ \protect \normalfont
  (\ignorespaces \ref {eq:state_of_maximum_EE}\unskip \@@italiccorr )}} for
  $n\leq M$ and its generalizations for cases $n\geq M$}\BibitemShut {NoStop}%
\bibitem [{\citenamefont {Rigol}\ \emph {et~al.}(2008)\citenamefont {Rigol},
  \citenamefont {Dunjko},\ and\ \citenamefont
  {Olshanii}}]{rigol2008thermalization}%
  \BibitemOpen
  \bibfield  {author} {\bibinfo {author} {\bibfnamefont {M.}~\bibnamefont
  {Rigol}}, \bibinfo {author} {\bibfnamefont {V.}~\bibnamefont {Dunjko}}, \
  and\ \bibinfo {author} {\bibfnamefont {M.}~\bibnamefont {Olshanii}},\
  }\href@noop {} {\bibfield  {journal} {\bibinfo  {journal}
  {\href{http://www.nature.com/nature/journal/v452/n7189/full/nature06838.html}{Nature}}\
  }\textbf {\bibinfo {volume} {452}},\ \bibinfo {pages} {854} (\bibinfo {year}
  {2008})}\BibitemShut {NoStop}%
\bibitem [{\citenamefont {Santos}\ and\ \citenamefont
  {Rigol}(2010)}]{santos2010onset}%
  \BibitemOpen
  \bibfield  {author} {\bibinfo {author} {\bibfnamefont {L.~F.}\ \bibnamefont
  {Santos}}\ and\ \bibinfo {author} {\bibfnamefont {M.}~\bibnamefont {Rigol}},\
  }\href@noop {} {\bibfield  {journal} {\bibinfo  {journal}
  {\href{https://journals.aps.org/pre/abstract/10.1103/PhysRevE.81.036206}{Phys.
  Rev. E}}\ }\textbf {\bibinfo {volume} {81}},\ \bibinfo {pages} {036206}
  (\bibinfo {year} {2010})}\BibitemShut {NoStop}%
\bibitem [{\citenamefont {Santos}\ \emph {et~al.}(2012)\citenamefont {Santos},
  \citenamefont {Polkovnikov},\ and\ \citenamefont {Rigol}}]{santos2012weak}%
  \BibitemOpen
  \bibfield  {author} {\bibinfo {author} {\bibfnamefont {L.~F.}\ \bibnamefont
  {Santos}}, \bibinfo {author} {\bibfnamefont {A.}~\bibnamefont {Polkovnikov}},
  \ and\ \bibinfo {author} {\bibfnamefont {M.}~\bibnamefont {Rigol}},\
  }\href@noop {} {\bibfield  {journal} {\bibinfo  {journal}
  {\href{https://journals.aps.org/pre/abstract/10.1103/PhysRevE.86.010102}{Phys.
  Rev. E}}\ }\textbf {\bibinfo {volume} {86}},\ \bibinfo {pages} {010102(R)}
  (\bibinfo {year} {2012})}\BibitemShut {NoStop}%
\bibitem [{\citenamefont {Deutsch}\ \emph {et~al.}(2013)\citenamefont
  {Deutsch}, \citenamefont {Li},\ and\ \citenamefont
  {Sharma}}]{deutsch2013microscopic}%
  \BibitemOpen
  \bibfield  {author} {\bibinfo {author} {\bibfnamefont {J.~M.}\ \bibnamefont
  {Deutsch}}, \bibinfo {author} {\bibfnamefont {H.}~\bibnamefont {Li}}, \ and\
  \bibinfo {author} {\bibfnamefont {A.}~\bibnamefont {Sharma}},\ }\href@noop {}
  {\bibfield  {journal} {\bibinfo  {journal}
  {\href{https://journals.aps.org/pre/abstract/10.1103/PhysRevE.87.042135}{Phys.
  Rev. E}}\ }\textbf {\bibinfo {volume} {87}},\ \bibinfo {pages} {042135}
  (\bibinfo {year} {2013})}\BibitemShut {NoStop}%
\bibitem [{\citenamefont {Beugeling}\ \emph {et~al.}(2014)\citenamefont
  {Beugeling}, \citenamefont {Moessner},\ and\ \citenamefont
  {Haque}}]{beugeling2014finite}%
  \BibitemOpen
  \bibfield  {author} {\bibinfo {author} {\bibfnamefont {W.}~\bibnamefont
  {Beugeling}}, \bibinfo {author} {\bibfnamefont {R.}~\bibnamefont {Moessner}},
  \ and\ \bibinfo {author} {\bibfnamefont {M.}~\bibnamefont {Haque}},\
  }\href@noop {} {\bibfield  {journal} {\bibinfo  {journal}
  {\href{https://journals.aps.org/pre/abstract/10.1103/PhysRevE.89.042112}{Phys.
  Rev. E}}\ }\textbf {\bibinfo {volume} {89}},\ \bibinfo {pages} {042112}
  (\bibinfo {year} {2014})}\BibitemShut {NoStop}%
\bibitem [{\citenamefont {Alba}(2015)}]{alba2015eigenstate}%
  \BibitemOpen
  \bibfield  {author} {\bibinfo {author} {\bibfnamefont {V.}~\bibnamefont
  {Alba}},\ }\href@noop {} {\bibfield  {journal} {\bibinfo  {journal}
  {\href{https://journals.aps.org/prb/abstract/10.1103/PhysRevB.91.155123}{Phys.
  Rev. B}}\ }\textbf {\bibinfo {volume} {91}},\ \bibinfo {pages} {155123}
  (\bibinfo {year} {2015})}\BibitemShut {NoStop}%
\bibitem [{Note2()}]{Note2}%
  \BibitemOpen
  \bibinfo {note} {It does not matter much which particular values we choose,
  as long as $t,t',V,V'\neq 0$, the evolution is qualitatively the
  same.}\BibitemShut {Stop}%
\bibitem [{\citenamefont {Sugiura}\ and\ \citenamefont
  {Shimizu}(2012)}]{sugiura2012thermal}%
  \BibitemOpen
  \bibfield  {author} {\bibinfo {author} {\bibfnamefont {S.}~\bibnamefont
  {Sugiura}}\ and\ \bibinfo {author} {\bibfnamefont {A.}~\bibnamefont
  {Shimizu}},\ }\href@noop {} {\bibfield  {journal} {\bibinfo  {journal}
  {\href{https://journals.aps.org/prl/abstract/10.1103/PhysRevLett.108.240401}{Phys.
  Rev. Lett.}}\ }\textbf {\bibinfo {volume} {108}},\ \bibinfo {pages} {240401}
  (\bibinfo {year} {2012})}\BibitemShut {NoStop}%
\bibitem [{\citenamefont {Sugiura}\ and\ \citenamefont
  {Shimizu}(2013)}]{sugiura2013canonical}%
  \BibitemOpen
  \bibfield  {author} {\bibinfo {author} {\bibfnamefont {S.}~\bibnamefont
  {Sugiura}}\ and\ \bibinfo {author} {\bibfnamefont {A.}~\bibnamefont
  {Shimizu}},\ }\href@noop {} {\bibfield  {journal} {\bibinfo  {journal}
  {\href{https://journals.aps.org/prl/abstract/10.1103/PhysRevLett.111.010401}{Phys.
  Rev. Lett.}}\ }\textbf {\bibinfo {volume} {111}},\ \bibinfo {pages} {010401}
  (\bibinfo {year} {2013})}\BibitemShut {NoStop}%
\bibitem [{\citenamefont {Lubkin}(1978)}]{lubkin1978entropy}%
  \BibitemOpen
  \bibfield  {author} {\bibinfo {author} {\bibfnamefont {E.}~\bibnamefont
  {Lubkin}},\ }\href@noop {} {\bibfield  {journal} {\bibinfo  {journal}
  {\href{https://aip.scitation.org/doi/abs/10.1063/1.523763}{J. Math. Phys.}}\
  }\textbf {\bibinfo {volume} {19}},\ \bibinfo {pages} {1028} (\bibinfo {year}
  {1978})}\BibitemShut {NoStop}%
\bibitem [{\citenamefont {Lloyd}\ and\ \citenamefont
  {Pagels}(1988)}]{lloyd1988complexity}%
  \BibitemOpen
  \bibfield  {author} {\bibinfo {author} {\bibfnamefont {S.}~\bibnamefont
  {Lloyd}}\ and\ \bibinfo {author} {\bibfnamefont {H.}~\bibnamefont {Pagels}},\
  }\href@noop {} {\bibfield  {journal} {\bibinfo  {journal}
  {\href{https://www.sciencedirect.com/science/article/pii/0003491688900942}{Ann.
  Phys.}}\ }\textbf {\bibinfo {volume} {188}},\ \bibinfo {pages} {186}
  (\bibinfo {year} {1988})}\BibitemShut {NoStop}%
\bibitem [{\citenamefont {Dahlsten}\ \emph {et~al.}(2014)\citenamefont
  {Dahlsten}, \citenamefont {Lupo}, \citenamefont {Mancini},\ and\
  \citenamefont {Serafini}}]{dahlsten2014entanglement}%
  \BibitemOpen
  \bibfield  {author} {\bibinfo {author} {\bibfnamefont {O.~C.}\ \bibnamefont
  {Dahlsten}}, \bibinfo {author} {\bibfnamefont {C.}~\bibnamefont {Lupo}},
  \bibinfo {author} {\bibfnamefont {S.}~\bibnamefont {Mancini}}, \ and\
  \bibinfo {author} {\bibfnamefont {A.}~\bibnamefont {Serafini}},\ }\href@noop
  {} {\bibfield  {journal} {\bibinfo  {journal}
  {\href{https://iopscience.iop.org/article/10.1088/1751-8113/47/36/363001/meta}{J.
  Phys. A: Math. Theor.}}\ }\textbf {\bibinfo {volume} {47}},\ \bibinfo {pages}
  {363001} (\bibinfo {year} {2014})}\BibitemShut {NoStop}%
\bibitem [{\citenamefont {Browaeys}\ \emph {et~al.}(2016)\citenamefont
  {Browaeys}, \citenamefont {Barredo},\ and\ \citenamefont
  {Lahaye}}]{browaeys2016experimental}%
  \BibitemOpen
  \bibfield  {author} {\bibinfo {author} {\bibfnamefont {A.}~\bibnamefont
  {Browaeys}}, \bibinfo {author} {\bibfnamefont {D.}~\bibnamefont {Barredo}}, \
  and\ \bibinfo {author} {\bibfnamefont {T.}~\bibnamefont {Lahaye}},\
  }\href@noop {} {\bibfield  {journal} {\bibinfo  {journal}
  {\href{https://iopscience.iop.org/article/10.1088/0953-4075/49/15/152001/meta}{J.
  Phys. B}}\ }\textbf {\bibinfo {volume} {49}},\ \bibinfo {pages} {152001}
  (\bibinfo {year} {2016})}\BibitemShut {NoStop}%
\bibitem [{\citenamefont {Zhang}\ \emph {et~al.}(2017)\citenamefont {Zhang},
  \citenamefont {Pagano}, \citenamefont {Hess}, \citenamefont {Kyprianidis},
  \citenamefont {Becker}, \citenamefont {Kaplan}, \citenamefont {Gorshkov},
  \citenamefont {Gong},\ and\ \citenamefont {Monroe}}]{zhang2017observation}%
  \BibitemOpen
  \bibfield  {author} {\bibinfo {author} {\bibfnamefont {J.}~\bibnamefont
  {Zhang}}, \bibinfo {author} {\bibfnamefont {G.}~\bibnamefont {Pagano}},
  \bibinfo {author} {\bibfnamefont {P.~W.}\ \bibnamefont {Hess}}, \bibinfo
  {author} {\bibfnamefont {A.}~\bibnamefont {Kyprianidis}}, \bibinfo {author}
  {\bibfnamefont {P.}~\bibnamefont {Becker}}, \bibinfo {author} {\bibfnamefont
  {H.}~\bibnamefont {Kaplan}}, \bibinfo {author} {\bibfnamefont {A.~V.}\
  \bibnamefont {Gorshkov}}, \bibinfo {author} {\bibfnamefont {Z.-X.}\
  \bibnamefont {Gong}}, \ and\ \bibinfo {author} {\bibfnamefont
  {C.}~\bibnamefont {Monroe}},\ }\href@noop {} {\bibfield  {journal} {\bibinfo
  {journal} {\href{https://www.nature.com/articles/nature24654}{Nature}}\
  }\textbf {\bibinfo {volume} {551}},\ \bibinfo {pages} {601} (\bibinfo {year}
  {2017})}\BibitemShut {NoStop}%
\bibitem [{\citenamefont {Friis}\ \emph {et~al.}(2018)\citenamefont {Friis},
  \citenamefont {Marty}, \citenamefont {Maier}, \citenamefont {Hempel},
  \citenamefont {Holz{\"a}pfel}, \citenamefont {Jurcevic}, \citenamefont
  {Plenio}, \citenamefont {Huber}, \citenamefont {Roos}, \citenamefont {Blatt}
  \emph {et~al.}}]{friis2018observation}%
  \BibitemOpen
  \bibfield  {author} {\bibinfo {author} {\bibfnamefont {N.}~\bibnamefont
  {Friis}}, \bibinfo {author} {\bibfnamefont {O.}~\bibnamefont {Marty}},
  \bibinfo {author} {\bibfnamefont {C.}~\bibnamefont {Maier}}, \bibinfo
  {author} {\bibfnamefont {C.}~\bibnamefont {Hempel}}, \bibinfo {author}
  {\bibfnamefont {M.}~\bibnamefont {Holz{\"a}pfel}}, \bibinfo {author}
  {\bibfnamefont {P.}~\bibnamefont {Jurcevic}}, \bibinfo {author}
  {\bibfnamefont {M.~B.}\ \bibnamefont {Plenio}}, \bibinfo {author}
  {\bibfnamefont {M.}~\bibnamefont {Huber}}, \bibinfo {author} {\bibfnamefont
  {C.}~\bibnamefont {Roos}}, \bibinfo {author} {\bibfnamefont {R.}~\bibnamefont
  {Blatt}},  \emph {et~al.},\ }\href@noop {} {\bibfield  {journal} {\bibinfo
  {journal}
  {\href{https://journals.aps.org/prx/abstract/10.1103/PhysRevX.8.021012}{Phys.
  Rev. X}}\ }\textbf {\bibinfo {volume} {8}},\ \bibinfo {pages} {021012}
  (\bibinfo {year} {2018})}\BibitemShut {NoStop}%
\bibitem [{\citenamefont {Fitzpatrick}\ \emph {et~al.}(2017)\citenamefont
  {Fitzpatrick}, \citenamefont {Sundaresan}, \citenamefont {Li}, \citenamefont
  {Koch},\ and\ \citenamefont {Houck}}]{fitzpatrick2017observation}%
  \BibitemOpen
  \bibfield  {author} {\bibinfo {author} {\bibfnamefont {M.}~\bibnamefont
  {Fitzpatrick}}, \bibinfo {author} {\bibfnamefont {N.~M.}\ \bibnamefont
  {Sundaresan}}, \bibinfo {author} {\bibfnamefont {A.~C.~Y.}\ \bibnamefont
  {Li}}, \bibinfo {author} {\bibfnamefont {J.}~\bibnamefont {Koch}}, \ and\
  \bibinfo {author} {\bibfnamefont {A.~A.}\ \bibnamefont {Houck}},\ }\href@noop
  {} {\bibfield  {journal} {\bibinfo  {journal}
  {\href{https://journals.aps.org/prx/abstract/10.1103/PhysRevX.8.021012}{Phys.
  Rev. X}}\ }\textbf {\bibinfo {volume} {7}},\ \bibinfo {pages} {011016}
  (\bibinfo {year} {2017})}\BibitemShut {NoStop}%
\bibitem [{\citenamefont {Gambetta}\ \emph {et~al.}(2017)\citenamefont
  {Gambetta}, \citenamefont {Chow},\ and\ \citenamefont
  {Steffen}}]{gambetta2017building}%
  \BibitemOpen
  \bibfield  {author} {\bibinfo {author} {\bibfnamefont {J.~M.}\ \bibnamefont
  {Gambetta}}, \bibinfo {author} {\bibfnamefont {J.~M.}\ \bibnamefont {Chow}},
  \ and\ \bibinfo {author} {\bibfnamefont {M.}~\bibnamefont {Steffen}},\
  }\href@noop {} {\bibfield  {journal} {\bibinfo  {journal}
  {\href{https://www.nature.com/articles/s41534-016-0004-0}{npj Quantum Inf.}}\
  }\textbf {\bibinfo {volume} {3}},\ \bibinfo {pages} {2} (\bibinfo {year}
  {2017})}\BibitemShut {NoStop}%
\bibitem [{\citenamefont {Xu}\ \emph {et~al.}(2018)\citenamefont {Xu},
  \citenamefont {Chen}, \citenamefont {Zeng}, \citenamefont {Zhang},
  \citenamefont {Song}, \citenamefont {Liu}, \citenamefont {Guo}, \citenamefont
  {Zhang}, \citenamefont {Xu}, \citenamefont {Deng} \emph
  {et~al.}}]{xu2018emulating}%
  \BibitemOpen
  \bibfield  {author} {\bibinfo {author} {\bibfnamefont {K.}~\bibnamefont
  {Xu}}, \bibinfo {author} {\bibfnamefont {J.-J.}\ \bibnamefont {Chen}},
  \bibinfo {author} {\bibfnamefont {Y.}~\bibnamefont {Zeng}}, \bibinfo {author}
  {\bibfnamefont {Y.-R.}\ \bibnamefont {Zhang}}, \bibinfo {author}
  {\bibfnamefont {C.}~\bibnamefont {Song}}, \bibinfo {author} {\bibfnamefont
  {W.}~\bibnamefont {Liu}}, \bibinfo {author} {\bibfnamefont {Q.}~\bibnamefont
  {Guo}}, \bibinfo {author} {\bibfnamefont {P.}~\bibnamefont {Zhang}}, \bibinfo
  {author} {\bibfnamefont {D.}~\bibnamefont {Xu}}, \bibinfo {author}
  {\bibfnamefont {H.}~\bibnamefont {Deng}},  \emph {et~al.},\ }\href@noop {}
  {\bibfield  {journal} {\bibinfo  {journal}
  {\href{https://journals.aps.org/prl/abstract/10.1103/PhysRevLett.120.050507}{Phys.
  Rev. Lett.}}\ }\textbf {\bibinfo {volume} {120}},\ \bibinfo {pages} {050507}
  (\bibinfo {year} {2018})}\BibitemShut {NoStop}%
\bibitem [{\citenamefont {Neill}\ \emph {et~al.}(2018)\citenamefont {Neill},
  \citenamefont {Roushan}, \citenamefont {Kechedzhi}, \citenamefont {Boixo},
  \citenamefont {Isakov}, \citenamefont {Smelyanskiy}, \citenamefont {Megrant},
  \citenamefont {Chiaro}, \citenamefont {Dunsworth}, \citenamefont {Arya} \emph
  {et~al.}}]{neill2018blueprint}%
  \BibitemOpen
  \bibfield  {author} {\bibinfo {author} {\bibfnamefont {C.}~\bibnamefont
  {Neill}}, \bibinfo {author} {\bibfnamefont {P.}~\bibnamefont {Roushan}},
  \bibinfo {author} {\bibfnamefont {K.}~\bibnamefont {Kechedzhi}}, \bibinfo
  {author} {\bibfnamefont {S.}~\bibnamefont {Boixo}}, \bibinfo {author}
  {\bibfnamefont {S.~V.}\ \bibnamefont {Isakov}}, \bibinfo {author}
  {\bibfnamefont {V.}~\bibnamefont {Smelyanskiy}}, \bibinfo {author}
  {\bibfnamefont {A.}~\bibnamefont {Megrant}}, \bibinfo {author} {\bibfnamefont
  {B.}~\bibnamefont {Chiaro}}, \bibinfo {author} {\bibfnamefont
  {A.}~\bibnamefont {Dunsworth}}, \bibinfo {author} {\bibfnamefont
  {K.}~\bibnamefont {Arya}},  \emph {et~al.},\ }\href@noop {} {\bibfield
  {journal} {\bibinfo  {journal}
  {\href{https://science.sciencemag.org/content/360/6385/195.abstract}{Science}}\
  }\textbf {\bibinfo {volume} {360}},\ \bibinfo {pages} {195} (\bibinfo {year}
  {2018})}\BibitemShut {NoStop}%
\bibitem [{\citenamefont {Cerf}\ and\ \citenamefont
  {Adami}(1997)}]{cerf1997negative}%
  \BibitemOpen
  \bibfield  {author} {\bibinfo {author} {\bibfnamefont {N.~J.}\ \bibnamefont
  {Cerf}}\ and\ \bibinfo {author} {\bibfnamefont {C.}~\bibnamefont {Adami}},\
  }\href@noop {} {\bibfield  {journal} {\bibinfo  {journal}
  {\href{https://journals.aps.org/prl/abstract/10.1103/PhysRevLett.79.5194}{Phys.
  Rev. Lett.}}\ }\textbf {\bibinfo {volume} {79}},\ \bibinfo {pages} {5194}
  (\bibinfo {year} {1997})}\BibitemShut {NoStop}%
\bibitem [{\citenamefont {Preskill}(2016)}]{preskill2016quantum}%
  \BibitemOpen
  \bibfield  {author} {\bibinfo {author} {\bibfnamefont {J.}~\bibnamefont
  {Preskill}},\ }\href@noop {} {\bibfield  {journal} {\bibinfo  {journal}
  {\href{https://arxiv.org/pdf/1604.07450.pdf}{arXiv:1604.07450} [quant-ph]}\ }
  (\bibinfo {year} {2016})}\BibitemShut {NoStop}%
\bibitem [{\citenamefont {Winter}(2016)}]{winter2016tight}%
  \BibitemOpen
  \bibfield  {author} {\bibinfo {author} {\bibfnamefont {A.}~\bibnamefont
  {Winter}},\ }\href@noop {} {\bibfield  {journal} {\bibinfo  {journal}
  {\href{https://link.springer.com/content/pdf/10.1007/s00220-016-2609-8.pdf}{Commun.
  Math. Phys.}}\ }\textbf {\bibinfo {volume} {347}},\ \bibinfo {pages} {291}
  (\bibinfo {year} {2016})}\BibitemShut {NoStop}%
\bibitem [{\citenamefont {Witten}(2018)}]{witten2018mini}%
  \BibitemOpen
  \bibfield  {author} {\bibinfo {author} {\bibfnamefont {E.}~\bibnamefont
  {Witten}},\ }\href@noop {} {\bibfield  {journal} {\bibinfo  {journal}
  {\href{https://arxiv.org/abs/1805.11965}{arXiv:1805.11965} [hep-th]}\ }
  (\bibinfo {year} {2018})}\BibitemShut {NoStop}%
\bibitem [{\citenamefont {Um}\ \emph {et~al.}(2012)\citenamefont {Um},
  \citenamefont {Park},\ and\ \citenamefont {Hinrichsen}}]{um2012entanglement}%
  \BibitemOpen
  \bibfield  {author} {\bibinfo {author} {\bibfnamefont {J.}~\bibnamefont
  {Um}}, \bibinfo {author} {\bibfnamefont {H.}~\bibnamefont {Park}}, \ and\
  \bibinfo {author} {\bibfnamefont {H.}~\bibnamefont {Hinrichsen}},\
  }\href@noop {} {\bibfield  {journal} {\bibinfo  {journal}
  {\href{https://iopscience.iop.org/article/10.1088/1742-5468/2012/10/P10026}{J.
  Stat. Mech., Theory. Exp.}}\ }\textbf {\bibinfo {volume} {2012}},\ \bibinfo
  {pages} {P10026} (\bibinfo {year} {2012})}\BibitemShut {NoStop}%
\bibitem [{\citenamefont {De~Tomasi}\ \emph {et~al.}(2017)\citenamefont
  {De~Tomasi}, \citenamefont {Bera}, \citenamefont {Bardarson},\ and\
  \citenamefont {Pollmann}}]{de2017quantum}%
  \BibitemOpen
  \bibfield  {author} {\bibinfo {author} {\bibfnamefont {G.}~\bibnamefont
  {De~Tomasi}}, \bibinfo {author} {\bibfnamefont {S.}~\bibnamefont {Bera}},
  \bibinfo {author} {\bibfnamefont {J.~H.}\ \bibnamefont {Bardarson}}, \ and\
  \bibinfo {author} {\bibfnamefont {F.}~\bibnamefont {Pollmann}},\ }\href@noop
  {} {\bibfield  {journal} {\bibinfo  {journal}
  {\href{https://journals.aps.org/prl/abstract/10.1103/PhysRevLett.118.016804}{Phys.
  Rev. Lett.}}\ }\textbf {\bibinfo {volume} {118}},\ \bibinfo {pages} {016804}
  (\bibinfo {year} {2017})}\BibitemShut {NoStop}%
\bibitem [{\citenamefont {Ohya}\ and\ \citenamefont
  {Watanabe}(2010)}]{ohya2010quantum}%
  \BibitemOpen
  \bibfield  {author} {\bibinfo {author} {\bibfnamefont {M.}~\bibnamefont
  {Ohya}}\ and\ \bibinfo {author} {\bibfnamefont {N.}~\bibnamefont
  {Watanabe}},\ }\href@noop {} {\bibfield  {journal} {\bibinfo  {journal}
  {\href{https://www.mdpi.com/1099-4300/12/5/1194}{Entropy}}\ }\textbf
  {\bibinfo {volume} {12}},\ \bibinfo {pages} {1194} (\bibinfo {year}
  {2010})}\BibitemShut {NoStop}%
\bibitem [{\citenamefont {Dixon}\ \emph {et~al.}(2012)\citenamefont {Dixon},
  \citenamefont {Howland}, \citenamefont {Schneeloch},\ and\ \citenamefont
  {Howell}}]{dixon2012quantum}%
  \BibitemOpen
  \bibfield  {author} {\bibinfo {author} {\bibfnamefont {P.~B.}\ \bibnamefont
  {Dixon}}, \bibinfo {author} {\bibfnamefont {G.~A.}\ \bibnamefont {Howland}},
  \bibinfo {author} {\bibfnamefont {J.}~\bibnamefont {Schneeloch}}, \ and\
  \bibinfo {author} {\bibfnamefont {J.~C.}\ \bibnamefont {Howell}},\
  }\href@noop {} {\bibfield  {journal} {\bibinfo  {journal}
  {\href{https://journals.aps.org/prl/abstract/10.1103/PhysRevLett.108.143603}{Phys.
  Rev. Lett.}}\ }\textbf {\bibinfo {volume} {108}},\ \bibinfo {pages} {143603}
  (\bibinfo {year} {2012})}\BibitemShut {NoStop}%
\bibitem [{\citenamefont {Wilde}(2017)}]{wilde2017}%
  \BibitemOpen
  \bibfield  {author} {\bibinfo {author} {\bibfnamefont {M.~M.}\ \bibnamefont
  {Wilde}},\ }\enquote {\bibinfo {title}
  {\href{https://arxiv.org/abs/1106.1445}{From Classical to Quantum Shannon
  Theory}},}\ in\ \href@noop {} {\emph {\bibinfo {booktitle} {Quantum
  Information Theory}}}\ (\bibinfo  {publisher} {Cambridge University Press},\
  \bibinfo {year} {2017})\ pp.\ \bibinfo {pages} {xi--xii},\ \bibinfo {edition}
  {2nd}\ ed.\BibitemShut {Stop}%
\bibitem [{\citenamefont {T{\"u}rkpen{\c{c}}e}\ \emph
  {et~al.}(2017)\citenamefont {T{\"u}rkpen{\c{c}}e}, \citenamefont
  {Ak{\i}nc{\i}},\ and\ \citenamefont {{\c{S}}eker}}]{turkpencce2017quantum}%
  \BibitemOpen
  \bibfield  {author} {\bibinfo {author} {\bibfnamefont {D.}~\bibnamefont
  {T{\"u}rkpen{\c{c}}e}}, \bibinfo {author} {\bibfnamefont {T.~{\c{C}}.}\
  \bibnamefont {Ak{\i}nc{\i}}}, \ and\ \bibinfo {author} {\bibfnamefont
  {S.}~\bibnamefont {{\c{S}}eker}},\ }\href@noop {} {\bibfield  {journal}
  {\bibinfo  {journal}
  {\href{https://arxiv.org/abs/1709.03276}{arXiv:1709.03276} [quant-ph]}\ }
  (\bibinfo {year} {2017})}\BibitemShut {NoStop}%
\bibitem [{\citenamefont {Datta}\ \emph {et~al.}(2018)\citenamefont {Datta},
  \citenamefont {Hirche},\ and\ \citenamefont {Winter}}]{datta2018convexity}%
  \BibitemOpen
  \bibfield  {author} {\bibinfo {author} {\bibfnamefont {N.}~\bibnamefont
  {Datta}}, \bibinfo {author} {\bibfnamefont {C.}~\bibnamefont {Hirche}}, \
  and\ \bibinfo {author} {\bibfnamefont {A.}~\bibnamefont {Winter}},\
  }\href@noop {} {\bibfield  {journal} {\bibinfo  {journal}
  {\href{https://arxiv.org/abs/1810.03644}{arXiv:1810.03644} [quant-ph]}\ }
  (\bibinfo {year} {2018})}\BibitemShut {NoStop}%
\bibitem [{\citenamefont {Koch-Janusz}\ and\ \citenamefont
  {Ringel}(2018)}]{koch2018mutual}%
  \BibitemOpen
  \bibfield  {author} {\bibinfo {author} {\bibfnamefont {M.}~\bibnamefont
  {Koch-Janusz}}\ and\ \bibinfo {author} {\bibfnamefont {Z.}~\bibnamefont
  {Ringel}},\ }\href@noop {} {\bibfield  {journal} {\bibinfo  {journal}
  {\href{https://www.nature.com/articles/s41567-018-0081-4}{Nat. Phys.}}\
  }\textbf {\bibinfo {volume} {14}},\ \bibinfo {pages} {578} (\bibinfo {year}
  {2018})}\BibitemShut {NoStop}%
\bibitem [{\citenamefont {Fannes}\ and\ \citenamefont
  {Van~Ryn}(2012)}]{fannes2012connecting}%
  \BibitemOpen
  \bibfield  {author} {\bibinfo {author} {\bibfnamefont {M.}~\bibnamefont
  {Fannes}}\ and\ \bibinfo {author} {\bibfnamefont {N.}~\bibnamefont
  {Van~Ryn}},\ }\href@noop {} {\bibfield  {journal} {\bibinfo  {journal}
  {\href{https://iopscience.iop.org/article/10.1088/1751-8113/45/38/385003/meta}{J.
  Phys. A: Math. Theor.}}\ }\textbf {\bibinfo {volume} {45}},\ \bibinfo {pages}
  {385003} (\bibinfo {year} {2012})}\BibitemShut {NoStop}%
\end{thebibliography}%
\end{document}